\documentclass[pra,twocolumn,aps,amssymb,showpacs,superscriptaddress]{revtex4-1}

\usepackage{epsfig}
\usepackage{graphicx}
\usepackage{amsmath}
\usepackage{amssymb}
\usepackage{amsmath,amssymb}
\usepackage{enumerate}
\usepackage{hyperref}
\usepackage[normalem]{ulem}
\usepackage{cancel}
\usepackage{float}
\usepackage{comment}
\usepackage{bm}
\usepackage{color}
\definecolor{rosso}{rgb}{1,0,0}
\definecolor{verde}{rgb}{0,1,0}
\definecolor{blue}{rgb}{0,0,1}
\definecolor{verdescuro}{rgb}{0,0.5,0.5}
\definecolor{rossoscuro}{rgb}{0.7,0.3,0}
\definecolor{bluscuro}{rgb}{0.3,0,0.7}
\definecolor{magenta}{rgb}{1,0,1}

\newcommand{\om}{{\omega}}

\newcommand{\vQ}{{\bf Q}}

\newcommand{\vq}{{\bf q}}

\newcommand{\vk}{{\bf k}}
\newcommand{\vj}{{\mathbf j}}

\newcommand\beq{\begin{equation}}
\newcommand\eeq{\end{equation}}
\newcommand{\bea}{\begin{eqnarray}}
\newcommand{\eea}{\end{eqnarray}}

\begin{document}

\title{Critical current throughout the BCS-BEC crossover \\ with the inclusion of pairing fluctuations}

\author{L. Pisani}
\affiliation{School of Science and Technology, Physics Division, Universit\`{a} di Camerino, 62032 Camerino (MC), Italy}
\author{V. Piselli}
\affiliation{CNR-INO, Istituto Nazionale di Ottica, Sede di Firenze, 50125 (FI), Italy}
\author{G. Calvanese Strinati}
\email{giancarlo.strinati@unicam.it}
\affiliation{School of Science and Technology, Physics Division, Universit\`{a} di Camerino, 62032 Camerino (MC), Italy}
\affiliation{CNR-INO, Istituto Nazionale di Ottica, Sede di Firenze, 50125 (FI), Italy}


\begin{abstract}
The present work aims at providing a systematic analysis of the {\it current density versus momentum\/} characteristics for a fermionic superfluid throughout the BCS-BEC crossover,
even in the fully homogeneous case. 
At low temperatures, where pairing fluctuations are not strong enough to invalidate a quasi-particle approach, a sharp  threshold for the inception of a back-flow current is found, which sets the onset of dissipation and identifies the critical momentum
according to Landau. 
This momentum is seen to smoothly evolve from the BCS to the BEC regimes, whereby a single expression for the single-particle current density that includes pairing fluctuations enables us to incorporate on equal footing two quite distinct dissipative mechanisms, namely, pair-breaking and phonon excitations in the two sides of the BCS-BEC crossover, respectively. 
At finite temperature, where thermal fluctuations broaden the excitation spectrum and make the dissipative (kinetic and thermal) mechanisms intertwined with each other, 
an alternative criterion due to Bardeen is instead employed to signal the loss of superfluid behavior.
In this way, detailed comparison with available experimental data in linear and annular geometries is significantly improved with respect to previous approaches, thereby demonstrating the crucial role played by quantum fluctuations in renormalizing the single-particle excitation spectrum.

\end{abstract}

\maketitle

\section{Introduction} 
\label{sec:introduction}

The concepts of the critical current of a superconductor \cite{Tinkham-1996} and the critical velocity of a superfluid \cite{LL-1980} are intimately related. 
In both cases, the breaking of dissipation-less flow occurs when the imparted kinetic energy gives rise to quasi-particle excitations with zero energy. 
For superconductors, this shows up in the phenomenon known as ``gapless superconductivity'' \cite{Fulde-1965,Parmenter-1965}
after the seminal work by Abrikosov and Gor'kov on the effect of impurities in a superconductor \cite{Abrikosov-1960}, which was later generalized by Maki \cite{Parks-1969} 
to the case of a time-reversal symmetry breaking (de-pairing or pair-breaking) agent \cite{Fulde-1963}.
For superfluids, it corresponds to the criterion for the onset of viscous flow originally conceived by Landau for $^{4}$He \cite{LL-1980}. 
To the extent that the properties of superconductors and superfluids crucially depend on the details of the underlying energy spectrum, a detailed analysis of the onset of the dissipative flow should be able to reveal the microscopic mechanisms underlying this phenomenon.

The appropriate physical quantity for revealing the above features is the \emph{current response to an overall  momentum\/} imposed on the (superconducting/superfluid) system.
The behavior of this physical quantity is familiar in two paradigmatic cases, namely, the weakly-attractive Fermi gas and the weakly-repulsive Bose gas. 
The first one can be described in terms of the Bardeen-Cooper-Schrieffer (BCS) theory of conventional superconductors \cite{Bardeen-1962} (with its extension to dirty superconductors \cite{Kupri-1980}),
while the second one in terms of the Bogoliubov theory of weakly-interacting bosons \cite{Fetter-1972} that Bose-Einstein condense (BEC) at low temperature (with its extension to strongly-interacting bosons like $^{4}$He \cite{Kramer-1969}).
In both systems, kinetic and/or thermal dissipation occurs in the form of a back-flow current (which is somewhat analogous to that occurring in a normal Fermi system \cite{Nozieres-1997}), giving rise to density inhomogeneities which, in turn, 
lead to the formation of vortices \cite{Tinkham-1996} as well as of rotons \cite{Feynman-1956}.

With the advent of the ultra-cold atomic gases (and, specifically, of Fermi gases with attractive inter-particle interaction), these two paradigmatic (fermionic and bosonic) physical systems can be smoothly connected via Fano-Feshbach resonances, whereby these systems correspond to the limiting regimes of the so-called BCS-BEC crossover, with largely-overlapping Cooper pairs in the BCS regime smoothly evolving into tightly-bound composite bosons in the BEC regime.
(A recent comprehensive review on the BCS-BEC crossover can be found in Ref.~\cite{Physics-Reports-2018}.)
Commonly, the BCS-BEC crossover is spanned in terms of the dimensionless coupling $(k_{F} a_{F})^{-1}$, where $k_{F}=(3 \pi^{2} n)^{1/3}$ is the Fermi momentum with density $n$ and $a_{F}$ the scattering length of the two-fermion problem in vacuum.
This coupling ranges from $(k_{F}\, a_{F})^{-1} \lesssim -1$ in the weak-coupling (BCS) regime when $a_{F} < 0$, to $(k_{F}\, a_{F})^{-1} \gtrsim +1$ in the strong-coupling (BEC) regime when $a_{F} > 0$, passing through the unitary limit $(k_{F}\, a_{F})^{-1} = 0$ when $|a_{F}|$ diverges.

Experiments with ultra-cold Fermi gases, with their detailed control on the relevant parameters associated with the degrees of freedom of the system Hamiltonian,
have revived the interest in many key aspects of superconductivity (or, more generally, of fermionic superfluidity), including
the Abrikosov vortex lattice \cite{Zwierlein-2005,Simonucci-2015},
the Josephson effect \cite{Miller-2007,Spuntarelli-2007},
and the interaction of magnetic(-like) fields with fermion spins \cite{Zwierlein-2006,Partridge-2006,Pieri-2006}.
At the same time, these experiments have also stimulated more refined and advanced theoretical approaches addressing these aspects, inevitably promoting a valuable feedback in the theory of superconductors themselves.
The topic dealt with in the present article, about the current response to an overall momentum imposed on a (homogeneous) fermionic superfluid, makes no exception to this close connection between experiment and theory
which is feasible with ultra-cold Fermi gases.
In particular, two recent experiments have measured at low temperature throughout the BCS-BEC crossover,
both the critical velocity in a linear geometry where a weak barrier procedes through the superfluid in a circular motion \cite{Weimer-2015},
and the maximum quantized persistent current circulating in an annular geometry \cite{DelPace-2022}.
Both experiments were essentially aimed at identifying the value of the critical current (or velocity) for an ultra-cold Fermi gas, in a way that would resemble as closely as possible the Landau criterion envisioned theoretically
for liquid helium \cite{Landau-1941}.

In this context, the present article considers the smooth evolution of the current-vs-momentum characteristics from the BCS to the BEC limits of the BCS-BEC crossover at any temperature in the superfluid phase, 
resting on an approach recently developed in Ref.~\cite{Pisani-2023} to deal with the effects of pairing fluctuations in the presence of a supercurrent and nontrivial spatial constraints on equal footing (although in the present case 
the spatial constraint may at most correspond to a weak barrier in line with the setups of the experiments of Refs.~\cite{Weimer-2015,DelPace-2022} mentioned above).
This theoretical approach will enable us to demonstrate, in terms of a single theory, how the Landau critical velocity at zero temperature, originally introduced for a bosonic superfluid \cite{Landau-1941}, smoothly evolves into its counterpart in the BCS regime where, however, the presence of an underlying Fermi surface makes the value of the critical velocity different from that obtained by the Bardeen criterion originally introduced for a fermionic superfluid \cite{Bardeen-1962}.
In addition, the analysis of the current-vs-momentum characteristics obtained in this way will enable us to identify the critical current (or, else, the critical velocity) even at finite temperature in terms of the Bardeen criterion.

The results for the current-momentum response obtained by the present approach, which include the renormalization of the single-particle excitation spectrum due to pairing fluctuations, represents a definite improvement over those given by the conventional BCS-RPA (Random Phase Approximation) theory of superconductors, originally introduced by Anderson for conventional superconductors \cite{Anderson-1958} and later extended to the BCS-BEC crossover \cite{MPS-1998,Comb-2006,Spunta-2010}.

The main results obtained in this article are as follows:

\noindent
(i) The current-vs-momentum characteristics are obtained with the inclusion of pairing fluctuations beyond mean field for a homogeneous two-component Fermi gas with attractive inter-particle interaction, 
at any temperature in the superfluid phase and throughout the BCS-BEC crossover.
Although, in principle, the value of the critical current obtained in this way is equivalent to that due to an applied infinitesimal perturbation, in practice the presence of a non negligible impurity (in the form of a small barrier) embedded in an otherwise homogeneous superfluid has to be taken into account when simulating realistic experimental setups.
This will be done in terms of the theoretical mLPDA approach (with the acronym standing for modified Local Phase Density Approximation) that was 
recently developed in Ref.~\cite{Pisani-2023}, with the further consideration of the extended Gorkov-Melik-Barkhudarov (GMB) approach implemented  in Ref.~\cite{Pisani-2018-b}
that improves on the comparison with experimental data.

\noindent
(ii) A rather good comparison is retrieved in this way with the experimental data available at low temperature, specifically, from Ref.~\cite{Weimer-2015} for the critical velocity in a linear geometry and from Ref.~\cite{DelPace-2022} for the maximum value of the quantized velocity in an annular geometry.

\noindent
(iii) The concept of the \emph{intrinsic\/} critical current is further extended \emph{locally\/} inside a realistic barrier, and shown to account for the value of the critical current which is involved in the Josephson effect 
at finite temperature as obtained in Ref.~\cite{Pisani-2023}.

\noindent
(iv) Overall, this analysis enables us to assess how the Landau and Bardeen criteria manifest themselves in different experimental contexts and at the relevant temperature, when the coupling is varied across the BCS-BEC crossover.

The article is organized as follows. 
In Section~\ref{sec:intrinsic-current-BCS-and_BEC} the current-vs-momentum characteristics are considered for the simple cases of a weakly-attractive Fermi gas and of a weakly-repulsive Bose gas, that
represent the limiting cases occurring in the BCS-BEC crossover.
In Section~\ref{sec:intr} the inclusion of pairing fluctuations beyond mean field will allow us to obtain a continuous evolution of the current-vs-momentum characteristics for a homogeneous Fermi system that evolves from the BCS to the BEC regimes,
thus making the intrinsic critical current obtained in this way to be the theoretical benchmark for interpreting microscopically experiments and theoretical simulations.
A linear geometry is first considered to relate with the experimental results on the critical velocity of Ref.~\cite{Weimer-2015} across the BCS-BEC crossover 
as well as with the theoretical results  on the Josephson characteristics obtained  in Ref.~\cite{Pisani-2023}.
Section~\ref{sec:persistent-current} considers alternatively an annular geometry with quantized values of the superfluid velocity, for which comparison with the experimental data of Ref.~\cite{DelPace-2022} is possible.
Section~\ref{sec:conclude} gives our conclusions.

Finally, for the benefit of the readers Appendix~\ref{sec:Appendix-A} briefly summarizes previous theoretical results which are utilized for the specific purposes of the present work,
while Appendix~\ref{sec:Appendix-B} expands on the topics dealt with in Secs.~\ref{sec:bcsj} and \ref{sec:LB-intrinsic}.

\section{Intrinsic critical current of weakly-interacting gases}
\label{sec:intrinsic-current-BCS-and_BEC}

This Section briefly reviews the current-density response induced by an imposed momentum in two paradigmatic cases: 
The weakly-attractive Fermi gas treated within the BCS approximation and the weakly-repulsive Bose gas treated within the Bogoliubov approximation. 
In the Bogoliubov case, the \emph{Landau critical velocity\/} sets not only the dissipative threshold but also the onset of  the dynamical instability of the gas.
In the BCS case, on the other hand, the dissipative threshold does not coincide with the onset of the dynamical instability.
This is due to the presence of an underlying Fermi surface which introduces a second (thermodynamic) critical velocity, that can be referred to as the \emph{Bardeen critical velocity\/} after Bardeen who first considered it \cite{Bardeen-1962}.

The expressions of the current density utilized here for the Fermi and Bose gases can be obtained as limiting cases of a more general expression which spans the BCS-BEC crossover between the two (BCS and BEC) limits,
which will be extensively discussed in Sec.~\ref{sec:intr} below.
In addition, for the benefit of the readers Appendix~\ref{sec:Appendix-A} summarizes how this more general expression for the current was originally obtained in Ref.~\cite{Pisani-2023}.

\vspace{-0.2cm}
\subsection{Intrinsic critical current of a \\ \hspace{0.6cm} weakly-attractive Fermi gas} 
\label{sec:bcsj}

We first consider the superfluid current density of a weakly-attractive Fermi gas as a function of an imposed velocity, which we calculate at the mean-field (BCS) level. 
Although this topic has already been dealt with by a number of authors \cite{Bardeen-1962,Parks-1969,Fulde-1963,Hansen-1968} in the context of de-pairing currents and gapless superconductivity
we are here interested in its intimate relation with the system energy excitation spectrum and with the Landau criterion for superfluidity \cite{Khalatnikov-2000}.

At the mean-field level, the current density induced by an imposed momentum $\vq$ reads \cite{Piselli-2020}
\begin{equation}
\vj(\vq) = n {\vq \over m} + 2 \! \int \! {d\vk \over (2\pi)^3} \, {\vk \over m} \; f(E_+(\vk;\vq)) \, ,
\label{bcslandau}
\end{equation}
where $n$ is the (fermionic) particle density, $m$ the fermion mass, $E_+(\vk;\vq) = E(\vk;\vq)+{\vk \cdot \vq \over m}$ the Doppler-shifted BCS quasi-particle spectrum where
$E(\vk;\vq)= \sqrt{( { \mathbf{k}^2 \over 2m} -\mu + {\vq^2 \over 2m} )^2 + \Delta_{\vq}^2}$ with gap parameter $\Delta_{\vq}$ and chemical potential $\mu$, $f(\epsilon)=(e^{\epsilon \over k_{B}T}+1)^{-1}$
the Fermi distribution function, and $2$ the spin factor. [We set $\hbar = 1$ throughout.]

Here, as well as in the more general expressions (\ref{totcurrdens}) and (\ref{density-distribution-fluctuations}), the first term $n {\vq \over m}$ represents the total density $n$ moving uniformly in stationary equilibrium with an imposed momentum $q = mv$, 
while the second term corresponds to the depletion of the superfluid component in favor of the normal one due either to thermal or velocity effects.
This connects with the Landau two-fluid model, as discussed below for both fermions and bosons.

By symmetry consideration, the second term on the right-hand side of Eq.~(\ref{bcslandau}) is directed along -$\vq$ and has accordingly the microscopic interpretation of a \emph{back-flow current\/} \cite{Nozieres-1997}, 
which is set in by the produced excitations that eventually make the system to dissipate.
The mechanism of the back-flow is actually a widely ranging concept, which is invoked in the stability of vortices in superconductors \cite{Tinkham-1996} and in the creation of rotons in $^{4}$He 
(as proposed in the seminal work by Feynman \cite{Feynman-1956}). 
In this respect, the typical return flow of rotons is not a specific feature of a bosonic system like $^{4}$He, since it was also detected in a two-dimensional Fermi liquid  like $^{3}$He \cite{Godfrin-2012} 
as well as in a dipolar bosonic quantum gas \cite{Ferlaino-2018}.

For small momenta $q \ll k_F$, the expression (\ref{bcslandau}) recovers the Landau's two-fluid model whereby the back-flow term  implicitly defines the \emph{normal fluid density\/} at finite temperature \cite{LL-1980}
\beq
\rho_{n}(T) = 2 \! \int \! {d\vk \over (2\pi)^3}  \; { (\vk \cdot \hat{\vq})^2 \over m} \left(-{d f(E(\vk)) \over dE(\vk)} \right) 
\label{normal-density}
\eeq
where $E(\vk) = E(\vk;\vq=0)$.
In this way, the standard hydrodynamic relation $\vj(\vq) = n {\vq \over m}-\rho_n(T) {\vq \over m} = \rho_s(T){\vq \over m}$ is obtained, where $\rho_{s}(T) = n - \rho_{n}(T)$ is the \emph{superfluid density\/} at temperature $T$.
In the present work, we go beyond the linear regime of Eq.~(\ref{normal-density}) and focus especially on the non-linear super-critical effects that show up once the Landau dissipation threshold is reached \cite{Khalatnikov-2000}.

\begin{figure}[t]
\begin{center}
\includegraphics[width=8.85cm]{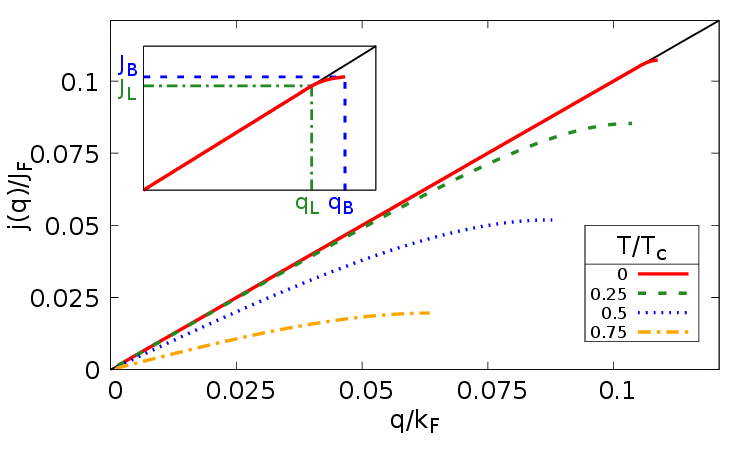}
	\caption{Momentum dependence of the current $j(q)$ for a weakly-attractive Fermi gas with coupling $(k_{F} a_{F})^{-1} = -1.0$, at several temperatures in the superfluid phase. 
                      Here, $q$ is in units of the Fermi momentum $k_{F} = \left(3 \pi^{2} n \right)^{1/3}$ and $j(q)$ in units of the Fermi current $J_{F} = n k_{F}/m$.
                      Main panel: At given temperature $T$ (here in units of the superfluid critical temperature $T_{c}$ calculated at the mean-filed level), the value $J_{B}$ of the critical current is reached at the maximum of each characteristic 
                      in agreement with the Bardeen criterion (see the text). 
	             Inset: At $T=0$, the onset of dissipation is defined by $q_{L}$ according to the Landau criterion, while at $q_{B}$ the stability is lost according to the Bardeen criterion.} 
\label{Figure-1}
\end{center}
\end{figure} 

Figure~\ref{Figure-1} shows the momentum dependence of the current density given by the expression (\ref{bcslandau}) at several temperatures.
In particular, at zero temperature the inset of Fig.~\ref{Figure-1}  shows that the current is proportional to the momentum $q$ and the back-flow term of Eq.~(\ref{bcslandau}) is zero as long as $q$ satisfies 
the Landau condition
\begin{multline}
E(\vk;\vq) - { k q \over m } \geq  0  \; \Rightarrow \; { q^2 \over  m } \le { q_{L}^2 \over  m } \equiv \sqrt{\Delta^2+\mu^2}-\mu 
\label{vL}
\end{multline}
which identifies the limiting (Landau) momentum $q_{L}$.
Past $q_{L}$, zero-energy excitations are produced in the system according to the Landau criterion, with the effect of decreasing the strength of the current in a gradual way up to a second critical (Bardeen) momentum $q_{B}$. 
At this point, the current reaches its maximum value which signals the limit of the thermodynamic stability of the system.
Past $q_{B}$, in fact, the dissipation becomes out of control in such a way that the thermodynamic equilibrium is lost and a new phase sets in.
Accordingly, at zero temperature the onset of dynamical dissipation occurring at the Landau momentum $q_{L}$ can be identified when the current-vs-momentum characteristic 
turns from a linear to a non-linear behavior, while the thermodynamic stability is lost at a second critical momentum $q_{B}$ when the characteristic attains its maximum value. 
A rigorous definition of $q_{B}$ will be given after the expression (\ref{bardcrit}) below when discussing the case of finite temperature.

At finite temperature, on the other hand, kinetic  and thermal dissipations are simultaneously present making the identification of $q_{L}$ from the current-vs-momentum characteristic no longer possible, 
since a sharp transition from a linear to a non-linear behavior is now lost (as shown in the main panel of Fig.~\ref{Figure-1}).
In this case, it is convenient to adopt a thermodynamic criterion for the stability of the system against the collapse of the superfluid phase due to dissipation as proposed by Bardeen \cite{Bardeen-1962}, 
which corresponds to the condition of the second derivative of the free energy $F$ being positive definite:
\beq
{\partial^2 F(q) \over \partial q^2} = {\partial j(q) \over \partial (q/m) } \ge 0 \, .
\label{bardcrit}
\eeq
Here, the equality corresponds to the maximum (or critical) current $J_{B}$ that occurs at the Bardeen momentum $q_{B}$. 
In this case, the superfuild density is defined directly as a dynamical response of the system in the form \cite{Bardeen-1962}
\beq
\rho_s(q)\equiv{\partial j(q) \over \partial (q/m)} \, ,
\label{superfluid-density}
\eeq
instead of being indirectly determined in terms of the normal density (\ref{normal-density}).
By this definition, $\rho_s(q)$ vanishes at the Bardeen momentum $q_{B}$ where $j(q)$ is maximum.
In addition, at zero temperatures $\rho_s(q)$ remains constant and equal to the particle density $n$ when $j(q)$ is linear in $q$ for $q \le q_{L}$, while it decreases to zero when $q_{L} < q \le q_{B}$.
These considerations will be further expanded in Appendix~\ref{sec:Appendix-B} in terms of the free energy at the mean-field level in the presence of a current.

In the related context of superconductivity, the maximum current $J_{c}$ is an intrinsic property of a superconductor and corresponds to the de-pairing current, 
whose knowledge is crucial in applications of thin and narrow superconducting films, like nanowire single-photon detectors and microwave kinetic inductance detectors \cite{Kubo-2020,Kunchur-2019,Clem-2012}.
Remaining in this related context, it may be instructive to relate the Landau threshold for viscous flow to the onset of gapless superconductivity. 
This can be done within the simplified context of mean-field theory, which is preparatory to the case of the BCS-BEC crossover (to be considered below) for which pairing fluctuations add to the picture.
It is known since the work of Gorkov and Abrikosov \cite{Abrikosov-1960} and its generalization due to Maki \cite{Parks-1969} that a dissipative agent (like magnetic impurities, a magnetic field, or a current) can make 
the \emph{excitation gap\/} in the single-particle spectrum to decrease to zero while maintaining the order parameter finite. 
Within mean-field (BCS) theory, when considering an imposed stationary flow with momentum $\vq$, the excitation spectrum is provided by the energy positions of the quasi-particle peaks in the single-particle spectral function
\beq
A(\vk,\om;\vq) = u^2_k \delta(\om-E_+(\vk;\vq)) + v^2_k \delta(\om+E_-(\vk;\vq))
\label{single-particle-spectral-function}
\eeq
where $E_{\pm}(\vk;\vq) = E(\vk;\vq)\pm{\vk \cdot \vq \over m}$ and $(u^2_k,v^2_k)$ are BCS coherence factors.
In this way, the total current density (\ref{bcslandau}) can then be rewritten as follows
\begin{equation}
\vj(\vq) = n {\vq \over m}+ \! \int \! {d\vk \over (2\pi)^3} \; {\vk \over m} \; n(\vk;\vq),
\label{totcurrdens}
\end{equation}
where we have introduced the \emph{density distribution function\/}
\beq
n(\vk;\vq)= 2 \int_{-\infty}^{+\infty} \!\!\! d\om f(\om) A(\vk,\om;\vq).
\label{density-distribution}
\eeq
In the simplest case of $T \rightarrow 0$ when thermal dissipation is absent, the onset of viscous flow given by the Landau criterion
\beq
E(\vk,\vq) - { k q \over m } = 0
\eeq
implies that the position of one of the two peaks in the spectral function (\ref{single-particle-spectral-function}) has shifted to zero energy, thereby signaling the closure of the excitation gap at some value of $\vk$ while the order parameter 
$\Delta_{\vq}$ remains finite. 
As a consequence, the Landau threshold at $q_{L}$ in the zero-temperature curve of Fig.\ref{Figure-1}  implies the closing of the excitation gap in the single-particle spectral function
(or, more conventionally, in the density of states \cite{Fulde-1965,Parks-1969}). 
This discussion will be relevant in the next Section when discussing the inclusion of pairing fluctuations. 

\begin{figure}[t]
\begin{center}
\includegraphics[width=9.1cm]{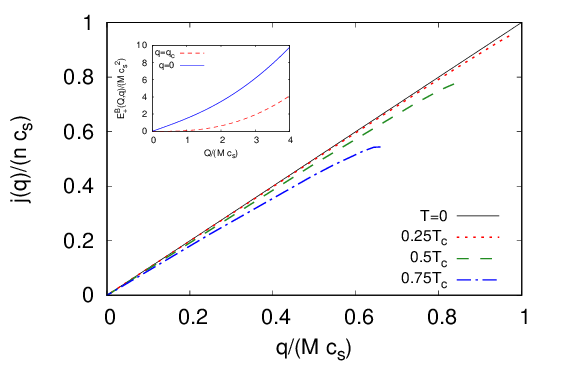}
	\caption{Momentum dependence of the current $j(q)$ for a weakly-repulsive Bose gas at several temperatures in the superfluid phase. 
                      At finite temperature, a maximum develops corresponding to the critical current $J_{c}$, which as $T\rightarrow 0$ coincides with the current flowing at the sound velocity.
	              The inset shows a comparison between the energy spectra of the Bose gas at rest and moving at the sound speed.}
\label{Figure-2}
\end{center}
\end{figure}

\vspace{-0.2cm}
\subsection{Intrinsic critical current of a \\ \hspace{0.6cm} a weakly-repulsive Bose gas}
\label{sec:wrbg}

The current density response of a weakly-interacting Bose gas moving with uniform momentum $\vq$ can similarly be studied \cite{Fetter-1972}.
In this case, the current density has an expression similar to Eq.~(\ref{bcslandau}) with suitable replacements, namely, 
\begin{equation}
\vj(\vq) = n {\vq \over M}+ \int \! {d\vQ \over (2\pi)^3} \; {\vQ \over M} \; b(E_+^B(\vQ;\vq))
\label{beclandau}
\end{equation}
where $n$ is the (bosonic) particle density, $M$ the boson mass, $E_{+}^{B}(\vQ;\vq) = E^{B}(\vQ,q)+{\vQ \cdot \vq \over M}$ the Doppler-shifted Bogoliubov quasi-particle spectrum with 
$E^{B}(\vQ,q) = \sqrt{\left({\vQ^2 \over 2M} +n_0(T,q)g\right)^{2} - \left(n_0(T,q)g\right)^2}$, $g$ the coupling constant, $n_0(T,q)$ the condensate density as determined consistently through the density equation for each imposed momentum q, and
$b(\epsilon)=(e^{\epsilon \over k_{B} T}-1)^{-1}$ is the Bose distribution function.

In Fig. \ref{Figure-2} the induced current is shown as a function of the impressed momentum $\vq$ for several temperatures below the critical temperature $T_{c}$ 
(which coincides with that of the non-interacting Bose gas, as consistently determined by the Bogoliubov theory), with the sound velocity $c_s=\sqrt{n_{0}(T=0,q=0) g \over M}$ being used for normalization in both axes.
At zero temperature, the straight line sharply drops to $-\infty$ as soon as the velocity exceeds $c_s$, thus signaling an instability of the system.
At finite temperature, the instability limit corresponds to the maximum of each curve, and represents the maximum (critical) current $J_{c}$ that can be sustained before a destructive flow sets in.

Beyond this point (and depending on the nature of the excitation spectrum of the Bose gas) a number of inhomogeneous phases have been considered in the literature, 
ranging from the condensation of rotons \cite{Iorda-1980} to the Bogoliubov-Cerenkov radiation \cite{Caru-2006,Baym-2012}, as well as to the recent observation of rotonic density modulations and supersolidity \cite{Modugno-2019}.
In the present work, we limit ourselves quite generally to considering the behavior of the current up to but not past the critical current.

\vspace{-0.2cm}
\section{Intrinsic critical current \\ throughout the BCS-BEC crossover}
\label{sec:intr}

In this Section, we perform a comprehensive study of the current-vs-momentum characteristics in the presence of pairing fluctuations for a fermionic superfluid spanning the BCS-BEC crossover.
By following the onset of dissipation signaled by these characteristics, we will show how the Landau critical velocity can be identified for \emph{all\/} couplings from the BCS to BEC limits.
This will be done by explicitly calculating a single measurable quantity, rather than by merely identifying the branches of the single-particle and two-particle excitation spectra 
corresponding to the lower velocity for given coupling, as one would instead do in a BCS-RPA approach \cite{Comb-2006,Spunta-2010}.

\vspace{-0.2cm}
\subsection{Current-vs-momentum characteristics \\ \hspace{0.6cm} in the presence of pairing fluctuations}
\label{pfchar}

The fermionic expression (\ref{totcurrdens}) for the total current density can quite generally be generalized to include pairing fluctuations, by adopting the following definition for the density distribution function
(cf. Ref.~\cite{Pisani-2023}  and Appendix~\ref{sec:Appendix-A} below):
\beq
n(\vk;\vq)= 2 k_{B}T \sum_n e^{i \om_n \eta} \, {\cal G}_{11}^{\mathrm{pf}}(\vk,\om_n;\vq) \, .
\label{density-distribution-fluctuations}
\eeq
Here, $\om_n = (2n+1)\pi k_{B}T$ ($n$ integer) is a fermionic Matsubara frequency, $\eta$ a positive infinitesimal, and ${\cal G}_{11}^{\mathrm{pf}}$ the ``normal'' single-particle 
Green's function obtained in the presence of a superfluid flow with momentum $\vq$.
In the following, the single-particle Green's function ${\cal G}_{11}^{\mathrm{pf}}$ will be calculated within the $t$-matrix approach in the presence of a super-current as developed in Ref.~\cite{Pisani-2023} 
(although when comparing with experimental data the refinements provided by the extended GMB approach of Ref.~\cite{Pisani-2018-b} will also be considered).
Like in the mean-field case (cf. Eq.~(\ref{bcslandau})), the current density given by Eqs.~(\ref{totcurrdens}) and (\ref{density-distribution-fluctuations}) is again made up of two terms,
a standard (classical) term proportional to the carriers velocity and a dissipative term identified as a back-flow current.

\begin{figure}[t]
\includegraphics[width=8.2cm]{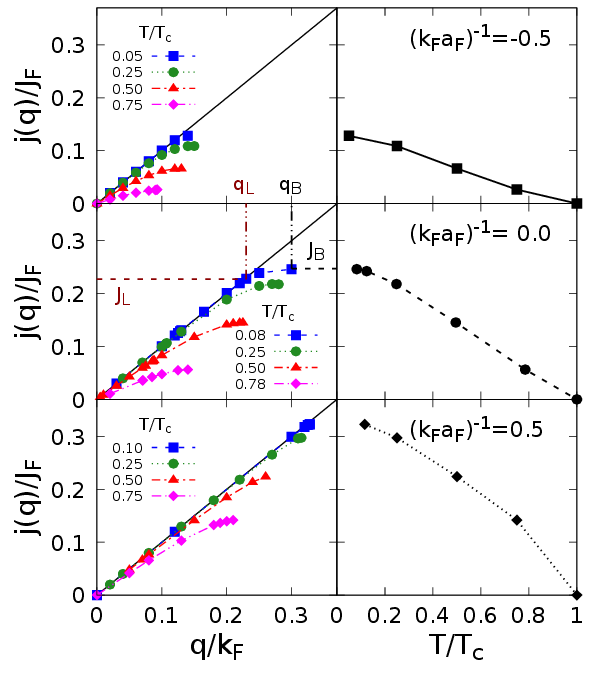}
\caption{The current-vs-momentum characteristics (normalized to $J_{F} = k_{F} n/m$) for three representative couplings across the BCS-BEC crossover and several temperatures in the superfluid phase, 
             as obtained by the $t$-matrix approach, are shown in the left panels. 
             For given coupling and temperature, the value of the critical current corresponds to the maximum of the curve according to the Bardeen criterion
             (where the temperature is now in units of the superfluid critical temperature $T_{c}$ calculated with the inclusion of pairing fluctuations).
             At zero temperature, both Landau ($J_{L}$) and Bardeen ($J_{B}$) critical currents can be identified, as shown at unitarity with their identifying labels in the left-central panel.
             In addition, the right panels report the temperature dependence of the  Bardeen critical current for given coupling.} 
\label{Figure-3}
\end{figure} 

Figure~\ref{Figure-3}  shows the current versus momentum characteristics obtained from Eqs.~(\ref{totcurrdens}) and (\ref{density-distribution-fluctuations}), for three representative couplings 
across the BCS-BEC crossover and several temperatures (left panels). 
At low-enough temperatures, the behavior is analogous to the mean-field case of Sec.~\ref{sec:bcsj}.
Accordingly, a threshold is found at a critical momentum, past which the current density is seen to deviate from a linear behavior and to enter a dissipative regime due to the onset of the back-flow current.
Below we will argue that this critical value corresponds to the Landau critical velocity $q_{L}/m$, where now the single-particle excitation branch of the spectrum is suitably renormalized by pairing fluctuations.

Similarly to the mean-field case of Sec.~\ref{sec:bcsj}, at any temperature the current density is seen to saturate at a maximum value, labelled in Fig.~\ref{Figure-3} 
by $J_{B}$ in correspondence to the Bardeen momentum $q_{B}$.
This value corresponds to the Bardeen criterion (\ref{bardcrit}) for the thermodynamic stability of the intrinsic current, such that past $q_{B}$ the system enters into a new phase.
The temperature dependence of the \emph{intrinsic\/} critical current $J_{B}$ for various couplings is reported in the right panels of Fig.~\ref{Figure-3}.
Note how the behavior of this temperature dependence confirms what was recently found in Ref.~\cite{Pisani-2023} (cf. Fig.~5 therein) for the critical current that can flow through a barrier,
whose temperature dependence changes from a convex to a concave behavior from the BCS to the BEC regime, passing through an essentially linear behavior at unitarity.
This linear temperature behavior of the critical current of a strongly-interacting fermionic superfluid is reminiscent of an analogous behavior of the critical velocity of $^{4}$He (as shown in Fig.~12 of Ref.~\cite{Varoquaux-2015}).

Note also from Fig.~\ref{Figure-3}  that the maximum of the characteristics about $q_{B}$ is much broader at unitarity than on either side of the crossover, 
and that the values of $J_{L}$ and $J_{B}$ remain close to each other in contrast to their respective momenta $q_{L}$ and $q_{B}$ which differ appreciably from each other.
Here, a clear (albeit not sharp) onset of dissipation is not restricted to zero temperature but extends, in practice, up to about $0.5 T_{c} \div 0.6 T{c}$, similarly to what was found in Sec.~\ref{sec:bcsj} for  a weakly-attractive Fermi gas.
This result is consistent with the fact that, in the broken-symmetry phase, the Landau damping due to thermal fluctuations becomes relevant above $0.3T_c \div 0.4T_c$, although 
the quasi-particle approximation (or, equivalently, the presence of a sharp energy spectrum on which the Landau criterion rests to begin with) ceases to be valid only above $0.5 \div 0.6T_c$ \cite{Pieri-2004}.
We shall return to this point below.

\vspace{-0.2cm}
\subsection{Landau and Bardeen intrinsic critical velocities at low temperature}
\label{sec:LB-intrinsic}

Quite generally, (quantum and thermal) pairing fluctuations are expected to give rise to a finite lifetime in the energy excitations, making the quasi-particle approximation much less reliable than in the mean-field case of Sec.~\ref{sec:bcsj}.
Nevertheless, by analyzing in detail the single- and two-particle excitation spectra in the presence of pairing fluctuations throughout the BCS-BEC crossover, in Ref.~\cite{Pieri-2004} it was shown that at low-enough temperature a quasi-particle picture is still applicable to the extent that the energy spectrum remains rather sharp.
In addition, it was shown that the single-particle spectral function $A(k,\om)$ possesses BCS-like features, with coherent peaks of quite small line-widths (cf. Fig.~10 of Ref.~\cite{Pieri-2004}) and dispersions showing a characteristic
back-bending (cf. Fig.~13 of Ref.~\cite{Pieri-2004}).
It was further shown that the degeneracy between the order parameter and the excitation gap occurring at the mean-field level is removed by pairing fluctuations. 
Specifically, a quantitative comparison between these two energy scales at low temperature was reported in Fig.~14 of Ref.~\cite{Pieri-2004}, showing a reduction of about $10\%$ of the excitation gap with respect to the order parameter at unitarity. 

Accordingly, as long as at low temperature the single-particle spectrum can be interpreted in terms of quasi-particle excitations with well-defined coherent peaks, 
one may assume a BCS-like expression like that given by Eq.~(\ref{single-particle-spectral-function}) for the single-particle spectral function to be valid throughout the BCS-BEC crossover
(although with renormalized values of the thermodynamic parameters appearing therein), thereby taking advantage of the arguments discussed in Sec.~\ref{sec:bcsj}.
This would imply that, even in the presence of pairing fluctuations, the onset of the non-linear behavior in $\vj(\vq)$ should reflect the closing of the excitation gap, thereby identifying the Landau critical threshold. 

\begin{figure}[t]
\begin{center}
\includegraphics[width=8.8cm]{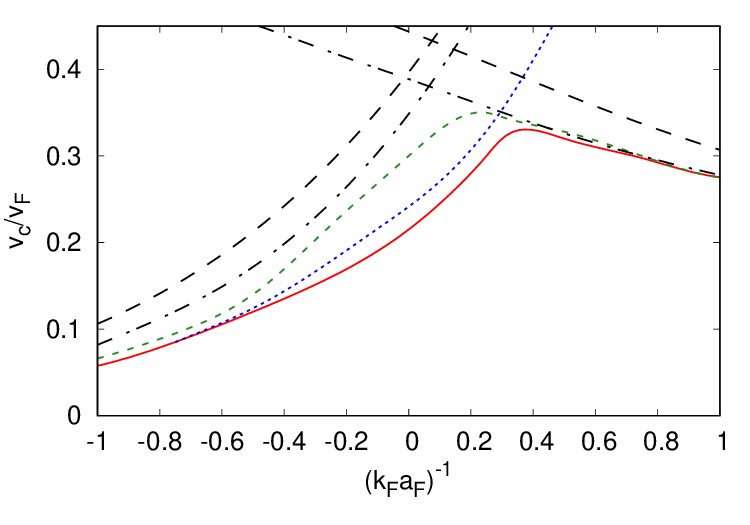}
\caption{Coupling dependence of the critical velocity $q_{c}/m$ at low temperature extracted from the characteristics $j(q)$ for various couplings.
	     Here, the red solid line represents the Landau velocity for which $q_{c} = q_{L}$ and the green dashed line the Bardeen velocity for which $q_{c} = q_{B}$, both obtained like in Fig.~\ref{Figure-3} 
	     by the $t$-matrix approach.
	     The blue dotted line corresponds to the critical velocity for pair-breaking excitations as obtained from the spectral function. 
             Long-dashed lines (dashed-dotted lines) correspond to pair-breaking (ascending branch) and sound (descending branch) critical velocities, with thermodynamic parameters that do not include (include) pairing fluctuations  
             \cite{Comb-2006,Spunta-2010} (see the text).}
\label{Figure-4}
\end{center}
\end{figure}

Figure~\ref{Figure-4} reports several curves for the \emph{critical velocity\/} $v_{c} = q_{c}/m$ throughout the BCS-BEC crossover at low temperature, with $q_{c}$ obtained from different approximations.
All curves contain an ``ascending'' branch associated with pair-breaking excitations in the BCS side of the crossover and a ``descending'' branch associated with phonon excitations in the BEC side of the crossover 
\cite{Spunta-2010}.

Initially, a BCS-RPA approach \cite{Anderson-1958} is adopted and extended throughout the BCS-BEC crossover \cite{MPS-1998,Spunta-2010,Comb-2006},
whereby the pair-breaking branch is obtained via the expression (\ref{vL}) for the Landau momentum $q_{L}$ and the phonon branch via the expression of the sound velocity at zero temperature reported in Ref.~\cite{MPS-1998},
with the thermodynamic parameters $\Delta$ and $\mu$ calculated either at the mean-field level (long-dashed lines \cite{Spunta-2010,Comb-2006}) or with the further inclusion of pairing fluctuations (dashed-dotted lines).
Both these pairs of lines represent an upper bound to the Landau critical velocity (cf. Fig.~8 of Ref.~\cite{Comb-2006} and Fig.~24 of Ref.~\cite{Spunta-2010}), and become strongly renormalized by dynamical many-body effects once properly included.

In this respect, the red solid line represents the value of the Landau critical velocity $v_{L} = q_{L}/m$ obtained at a low (but non-zero) temperature (T=0.1T$_c$) from the threshold of the non-linear behavior as identified in Fig.~\ref{Figure-3}.
Note here the strong suppression with respect to the previous BCS-RPA results (black long-dashed  \cite{Comb-2006,Spunta-2010} and dashed-dotted lines), which is due to dynamical effects that are not taken into account in those results, 
yielding a significant renormalization of the excitation gap and of the underlying Fermi surface. 
In addition, the short-dashed line of Fig.~\ref{Figure-4}  represents the velocity $q_{B}/m$ for which the current reaches its critical value, corresponding to the maximum value extracted from the left panels of 
Fig.~\ref{Figure-3}.

By taking advantage of the analytic considerations reported in Sec.~\ref{sec:bcsj}, the single-particle excitations at low temperature can be interpreted throughout the crossover in terms
of a quasi-particle approximation with an effective BCS-like single-particle spectral function.
Accordingly, following Ref.~\cite{Pieri-2004}, for given coupling we have obtained renormalized values of the order parameter and chemical potential by a BCS-like fit to the dispersion relation extracted from the single-particle spectral function, 
in terms of which we have calculated the Landau critical velocity for the ascending branch using the expression (\ref{vL}).
The result of this calculation corresponds to the (blue) dotted line of Fig.~\ref{Figure-4}.
Note how this line is quite close although not identical to the solid line therein, owing to a small broadening of the spectral line-shape associated with quantum fluctuations not captured by the quasi-particle approximation. 

Physically, the dotted line of Fig.~\ref{Figure-4}  represents the Landau velocity relative to the single-particle (pair-breaking) contribution to the excitation spectrum.
This is the reason why it tends to increase without bound  upon approaching the BEC regime, where it is expected to follow the behavior of the pair binding energy.
The solid line of Fig.~\ref{Figure-4}  instead switches smoothly from the ascending branch of pair-breaking excitations to the descending branch of phonon excitations,
with the maximum reached for coupling $(k_{F} a_{F})^{-1} = + 0.36$ (cf. Appendix~\ref{sec:Appendix-B} for for further considerations on this topic).
This smooth evolution of the \emph{intrinsic\/} critical velocity, from the ascending to the descending branches when the BCS-BEC crossover is spanned from the BCS to the BEC regimes, takes place directly from the expressions (\ref{totcurrdens}) and (\ref{density-distribution-fluctuations}) for the homogeneous case where pairing fluctuations are included. 
It is worth pointing out that this smooth evolution corresponds to a more realistic picture of the critical velocity, by removing the unphysical cusp present in the BCS-RPA approach 
(long-dashed and dashed-dotted black lines in  Fig.~\ref{Figure-4}) \cite{Spunta-2010, Comb-2006}.
At the mean-field level, on the other hand, a smooth evolution between the two branches can be recovered when the current is made to flow in the presence of a small (in the limit, infinitesimal) barrier that breaks translational invariance \cite{Spunta-2010,Piselli-2020}.

\begin{figure}[t]
\begin{center}
\includegraphics[width=8.7cm]{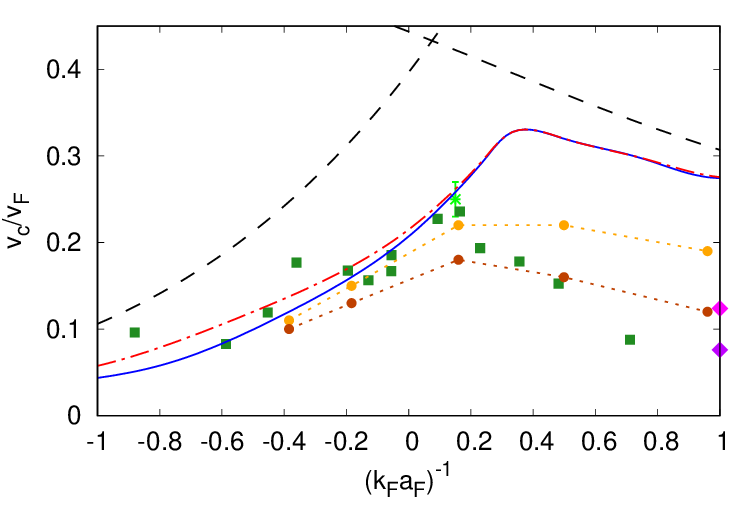}
\caption{Comparison with the experimental data for the Landau critical velocity at low temperature from Ref.~\cite{Miller-2007} (light green asterisk) and Ref.~\cite{Weimer-2015} (dark green squares). 
             Theoretical results obtained by the extended GMB approach of Ref.~\cite{Pisani-2018-b} in the homogeneous case are given by the blue solid line,
             and are compared with the results obtained by the $t$-matrix approach reproduced from Fig.~\ref{Figure-4} above (red dashed-dotted line).
	     Results obtained with the addition of a small Gaussian barrier (as specified in Ref.~\cite{Weimer-2015} for different number of atoms) are given by light and dark circles (joined by dotted lines for clarity)
	     The corresponding results obtained by the Gross-Pitaevskii equation for composite bosons with scattering length $0.6 a_{F}$ are reported as light and dark diamonds.
	      For completeness, the long-dashed line reproduces the BCS-RPA critical velocity also reported in Fig.~\ref{Figure-4} above.}
\label{Figure-5}
\end{center}
\end{figure}

In this respect, it may be instructive to consider the BEC regime of Eqs.~(\ref{totcurrdens}) and (\ref{density-distribution-fluctuations}), for which analytic expressions can be obtained (cf. Appendix~B of Ref.~\cite{Pisani-2023}). 
One obtains the following approximate expression for the density distribution function (\ref{density-distribution-fluctuations}):
\begin{multline}
        n(\vk,\vq) \simeq \frac{1}{4 \xi(\mathbf{k})^2} \left[2\Delta_\vq^2 \; + \right. \\
	\left. 2 \!\! \int \!\!\! \frac{d \mathbf{Q}}{(2 \pi)^3} k_{B} T \sum_{\nu} \! e^{i \Omega_{\nu} \eta} \, \Gamma_{11}(\mathbf{Q},\Omega_{\nu};\mathbf{q}) \left( 1+ \frac{\mathbf{k} \cdot \mathbf{Q} / m}{\xi(\mathbf{k})} \right) \right] 
\label{vkbec}
\end{multline}
where $\xi(\mathbf{k}) = \mathbf{k}^{2}/(2m) - \mu$, $\Omega_{\nu} = 2 \pi \nu k_{B} T$ ($\nu$ integer) is a bosonic Matsubara frequency, and $\Gamma_{11}(\mathbf{Q},\Omega_{\nu};\mathbf{q})$ is the diagonal element of
the particle-particle ladder in the presence of a super-current \cite{Pisani-2023}.
After integration over the fermionic variables, the total current density (\ref{totcurrdens}) becomes \cite{Pisani-2023}:
\begin{equation} 
\vj(\vq)- n {\vq \over m}  \simeq 2 \! \int \! \frac{d \mathbf{Q}}{(2 \pi)^3} \, \frac{\mathbf{Q}}{2 m} \, b \! \left(E_{+}^B(\mathbf{Q} ; \mathbf{q})\right)
\label{currdens-bec}
\end{equation}
where $E_{+}^B(\mathbf{Q})$ is given after Eq.~(\ref{beclandau}) and $2m = M$ is the mass of composite bosons that form in the BEC limit (to which an imposed momentum $2 \mathbf{q}$ is now associated).
It is thus evident that the fermionic current has become purely bosonic in nature, with the Landau threshold now set by the Bogoliubov sound velocity (cf. Sec.~\ref{sec:wrbg}). 

\vspace{-0.2cm}
\subsection{Comparison with the available experimental data for the critical velocity}

A direct comparison can be made at this point with the available experimental data for the Landau critical velocity $v_{L}$ obtained with an ultra-cold trapped Fermi gas,
initially at unitarity in Ref.~\cite{Miller-2007} and, more extensively, throughout the BCS-BEC crossover in Ref.~\cite{Weimer-2015}.
These data are reported in Fig.~\ref{Figure-5}  as a light green asterisk \cite{Miller-2007} and dark green squares \cite{Weimer-2015}, respectively.
To make the best possible comparison with these experimental data, we have improved on the $t$-matrix results shown previously in Fig.~\ref{Figure-4}  and implemented in the present context the extended GMB approach of Ref.~\cite{Pisani-2018-b}, which has recently proved to lead to a quite good comparison with experiments in several contexts \cite{Moritz-2022},\cite{Koehl-2023},\cite{Pisani-2023},\cite{Piselli-2023}.
[For the benefit of the readers, a concise summary of the extended GMB approach of Ref.~\cite{Pisani-2018-b} can be found in Appendix~\ref{sec:Appendix-A} below.]

The results of the extended GMB approach for the critical Landau velocity at low temperature in the homogeneous case are shown by the blue solid line of Fig.~\ref{Figure-5}.

To offer a direct comparison with the experimental results, Fig.~\ref{Figure-5}  also reproduces from Fig.~\ref{Figure-4}  
both the results obtained by the $t$-matrix approach (red dashed-dotted line) and the BCS-RPA critical velocity (long-dashed line).

A quite good agreement is found between the experimental data and the extended GMB approach for the ascending (pair-breaking) branch, at least for coupling values up to about $+0.2$ before
the descending (phononic) branch prevails and the maximum occurs,
with a slight (yet favorable) improvement on the BCS side of the crossover over and above the results of the $t$-matrix approach. 
It turns out, however, that the behavior of $v_{L}$ about and past this (intermediate) coupling regime is sensible to the presence even of the small barrier considered in the experiment of Ref.~\cite{Weimer-2015}. 
To quantify this effect, on top of the extended GMB approach we have considered a small Gaussian barrier like that utilized in the experiment of Ref.~\cite{Weimer-2015} (although of a different sign, which should be immaterial in the limit of infinitesimal barrier), with the results given by the light and dark circles in Fig.~\ref{Figure-5}.

Here, the presence of this barrier is dealt with the mLPDA approach of Ref.~\cite{Pisani-2018-b}, where pairing fluctuations are included on top of the original LPDA approach of Ref.~\cite{Simonucci-2014}.
[For the benefit of the readers, a concise summary of the LPDA approach of Ref.~\cite{Simonucci-2014} as well as of the mLPDA approach of Ref.~\cite{Pisani-2018-b} can be found in Appendix~\ref{sec:Appendix-A} below.]

The difference between these values stems from the experimental uncertainty in the number of atoms \cite{Weimer-2015}, which is reflected in the values of the Fermi momentum utilized to normalize the theoretical results 
(see below).

It is evident from Fig.~\ref{Figure-5}  that the presence of a small barrier has only minor effects on the ascending branch on the BCS side of the crossover up to (about) unitarity, but its effects become rather substantial on the BEC side of the crossover past unitarity when the descending branch is dominated by bosonic degrees of freedom and the underlying Fermi surface is lost.
This enhanced sensitivity to the presence of spatial inhomogeneities on the BEC side of the crossover is in line with what found in Ref.~\cite{Palest-2013} for the effects of random
impurities.

Yet, in Fig.~\ref{Figure-5}  discrepancies with the experimental data still remain deep in the BEC side of the crossover for $(k_{F} a_{F})^{-1} \approx 1$, 
even after having included the effects of a small barrier.
This discrepancy could be due to our theory recovering in this limit the (Born) value $2.0 a_{F}$ \cite{Pieri-2003} instead of the correct value $0.6 a_{F}$ \cite{Brodsky-2006} for the scattering length of composite bosons,
thereby overestimating the value of the speed of sound.
To clarify this point, we have calculated the Landau critical velocity at zero temperature for coupling $(k_{F} a_{F})^{-1} = 1.0$ in terms of the Gross-Pitaevskii equation, with the values $0.6 a_{F}$ for the bosonic scattering length.
The results of this additional calculation are shown in  Fig.~\ref{Figure-5}  by light and dark diamonds, respectively, where for consistency the presence of a small barrier is also taken into account as we did above.
One sees that, with these additions, in the BEC regime the theoretical results come considerably closer to the experimental data.

The residual discrepancy between theory and experiment, as far as the bosonic side of the descending branch in  Fig.~\ref{Figure-5}  is concerned, can be ascribed to
the method used in Ref.~\cite{Weimer-2015} to excite the gas of ultra-cold atoms through a circular stirring of a laser.
The same authors of Ref.~\cite{Weimer-2015} have proven in Ref.~\cite{Weimer-2016} that for a bosonic gas the additional centrifugal energy present in a circular stirring acts to lower significantly the value of the Landau critical velocity, 
which would instead be expected to coincide with the Bogoliubov sound velocity (see also Fig. 2(a) in Ref.~\cite{Weimer-2015}).
In contrast, the presence of an underlying Fermi surface on the BCS side of the crossover up to somewhat past unitarity makes the centrifugal energy negligible when compared
with the fermionic chemical potential.

We return, finally, to the values of the Fermi momentum mentioned above and utilized to normalize the theoretical results.
In Ref.~\cite{Miller-2007}, the experimental values of $v_L$ were expressed directly in terms of the \emph{local\/} Fermi velocity $v_F^{\mathrm{loc}}$ at the trap center, yielding the value $v_L/v_F^{\mathrm{loc}}=0.25$
reported in Fig.~\ref{Figure-5}  (light green asterisk).  
In Ref.~\cite{Weimer-2015}, on the other hand, the experimental values of $v_L$ were given in terms of the \emph{global\/} (trap) Fermi velocity $v_F^t$.
To estimate the corresponding local values of $v_F^{\mathrm{loc}}$ at the trap center, we have taken advantage of the procedure followed in Ref.~\cite{Weimer-2015} when converting 
the theoretical values for the speed of sound obtained in Ref.~\cite{Astra-2004} in the homogeneous case to the experimental trap geometry throughout the whole BCS-BEC crossover.
This procedure effectively amounts to multiplying the value of the sound velocity obtained in the homogeneous case by the factor $v_F^{\mathrm{loc}}/v_F^t$. 
We have used this procedure in reverse and extracted the geometrical factor $v_F^t/v_F^{\mathrm{loc}}$ specific to the trap settings of Ref.~\cite{Weimer-2015}, which has enabled 
us to convert the value of the experimental data of Ref.~\cite{Weimer-2015} for the Landau critical velocity in a way to compare them with our theoretical results for the homogeneous case. 

\vspace{-0.2cm}
\subsection{Josephson critical current interpreted as \\ \hspace{0.7cm} an intrinsic critical current inside the barrier}

\begin{figure}[t]
\begin{center}
\includegraphics[width=8.9cm]{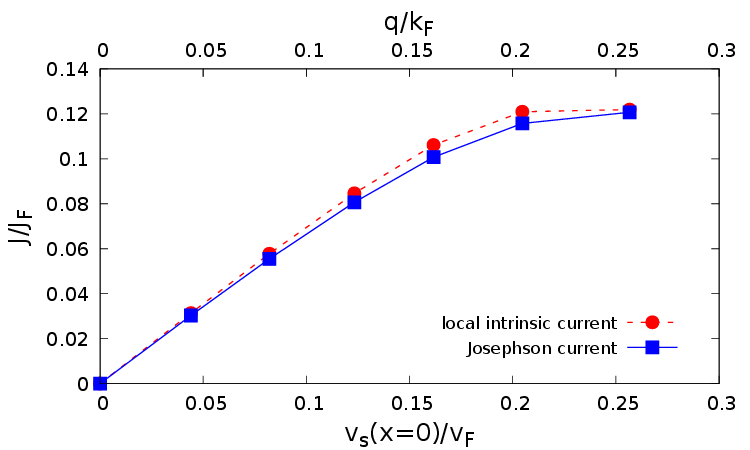}
	\caption{The Josephson characteristic, obtained at unitarity and $T/T_{c}=0.15$ for a Gaussian barrier of width $2.5 k_{F}^{-1}$ and height $V_{0}/E_{F}=0.1$, as expressed in terms of the local superfluid velocity $v_{s}(x=0)$
	             at the barrier center (filled squares), is compared with the intrinsic current obtained for a homogeneous system using the ``local'' thermodynamic parameters determined at the barrier center (filled circles).}
\label{Figure-6}
\end{center}
\end{figure}

The considerations made so far for the \emph{intrinsic\/} critical current (or critical velocity) have naturally direct applications to physical systems which are either homogeneous or slightly deviate from a homogeneous condition.
In this Section we apply the concept of critical current to the case of the Josephson effect for a fermionic superfluid flowing across a barrier of width larger than or comparable with the healing length of the bulk superfluid.
This condition is satisfied by most experiments with ultra-cold Fermi atomic gases, for which the healing length is of the order of the inter-particle spacing $k_F^{-1}$ in the relevant 
coupling range $-0.5 < (k_{F}a_{F})^{-1} < 1$ probed by the experiments, while the width of the barrier cannot be made as small \cite{Kwon-2020,DelPace-2021}.

In the Josephson effect, the maximum (critical) current that can flow across a barrier depends on the shape (height and width) of the barrier, besides coupling and temperature.
In Ref.~\cite{Pisani-2023} a systematic study of the Josephson characteristics was performed in terms of a theoretical (mLPDA)  approach that allows for the inclusion of pairing fluctuations in a non-trivial spatial geometry,
 thus complementing and extending the original LPDA approach of Ref.~\cite{Simonucci-2014}.
[For the benefit of the readers, a concise summary of the LPDA approach of Ref.~\cite{Simonucci-2014} as well as of the mLPDA approach of Ref.~\cite{Pisani-2018-b} can be found in Appendix~\ref{sec:Appendix-A} below.]

Here, we analyze the case examined in Figs.~2 and 3 of Ref.~\cite{Pisani-2023}, where a Gaussian barrier of width $2.5 k_{F}^{-1}$ and height $V_{0}/E_{F}=0.1$ was considered for temperature $T/T_{c}=0.15$ at unitarity.
It turns out that the maximum current sustained by the barrier is quantitatively related to the intrinsic critical current of a bulk superfluid with the same ``local'' 
thermodynamic conditions that develop at the center of the barrier.

This behavior is shown in Fig.~\ref{Figure-6}, where the current density as obtained by the methods of Ref.~\cite{Pisani-2023} is reported as a function of the
\emph{local\/} superfluid velocity $v_{s}(x=0)$ at the center of the barrier, and compared with the intrinsic current $\vj(\vq)$ where $|\vq|=m v_s$ as obtained from Eqs.~(\ref{totcurrdens}) and (\ref{density-distribution-fluctuations}) for a homogeneous superfluid with the same coupling and temperature, in which the local thermodynamic parameters $\Delta(x=0)$, $\mu(x=0)$, and $n(x=0)$ at the center of the barrier are used. 
From the good agreement between these two curves we are led to conclude that it is the dissipative mechanism of the back-flow current which develops inside the barrier 
to determine the value of the critical current for the Josephson junction.
Note how, in both cases, the maximum (critical) currents correspond to the Bardeen critical current discussed in Sec.~\ref{pfchar}.

\vspace{-0.2cm}
\section{Decay of persistent currents}
\label{sec:persistent-current}

A topic for which the concept of intrinsic critical current discussed above is especially relevant is that of the persistence of a super-current induced in a superfluid with a closed (annular) geometry and of the associated decay mechanisms, 
which are the hallmark of superfluidity in the first place.
In principle, a persistent current will continue indefinitely as long as the medium is superfluid.
In practice, persistent currents in superconducting materials (like, for instance, NbZr alloys) were estimated to flow over more than hundreds of years \cite{File-Mills-1963}.
Recently, this topic was taken over in the context of ultra-cold Fermi gases, for which a recent experiment \cite{DelPace-2022} using a phase-imprinting technique has detected
quantized circulations across the BCS-BEC crossover persisting up to a few seconds and identified its decay mechanism to take place via the emission of vortices \cite{Xhani-2023}.
In this Section, we examine the outcomes of this experiment in the light of the theoretical considerations made above in Sec.~\ref{sec:intr}.

\begin{figure}[t]
\begin{center}
\includegraphics[width=7.75cm]{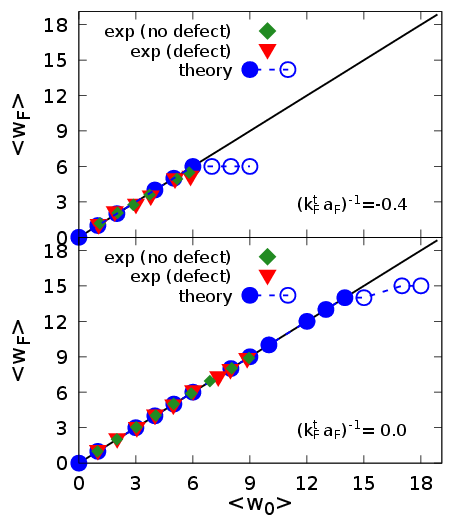}
 \caption{Experimental data for the final mean winding number $\langle w_{F} \rangle$ vs the impressed winding number $\langle w_{0} \rangle$ are compared with theoretical calculations 
               for couplings $(k_{F}a_{F})^{-1} = -0.4$ (top panel) and $(k_{F}a_{F})^{-1} = 0.0$ (bottom panel).
              Experimental data for a homogeneous superfluid (diamonds) are from Fig.~3(c) and those in presence of a point defect (triangles) from Fig.~4(e) of Ref.~\cite{DelPace-2022}.
              The results of the theoretical simulations at temperature $T=0.02 T_{F}$ (as described in detail in the text) are identified by filled dots up to the
              critical value of $\langle w_{F} \rangle$ and by empty dots past this critical value.}
\label{Figure-7}
\end{center} 
\end{figure}

Typically, along a circle of radius $R$ the velocity field $\mathbf{v} = \nabla \varphi(\mathbf{r}) / (2m)$ of the fermionic superfluid (where $\varphi(\mathbf{r})$ is the phase of the local gap parameter 
$\Delta(\mathbf{r}) = |\Delta(\mathbf{r})| e^{i\varphi(\mathbf{r})}$ and $m$ is the fermion mass) has a quantized circulation given by $\pi w / m$ where $w$ is an integer.
Correspondingly, the magnitude of $\mathbf{v}$ takes the quantized values $v = w / (2m R)$.
The integer $w$ is referred to as a \emph{winding number\/} just because it counts the number of oscillations of the phase of the order parameter along the circulation.
It can be measured by interferometric techniques.
After an initial overall phase difference $2\pi w_{0}$ has been imprinted on the superfluid, the system is observed to stabilize
for a time of the order of tenths of a second and its final (mean) winding number $\langle w_{F} \rangle$ is measured. 
Experimentally, mean winding numbers (both for $w_{0}$ and $w_{F}$) are considered because the same imprinting procedure is repeated several times, such that the ensuing results are suitably averaged out.

Figure~\ref{Figure-7}  reports the experimental data from Ref.~\cite{DelPace-2022} for the mean winding number $\langle w_{F} \rangle$ at low temperature 
for the couplings $(k_{F}a_{F})^{-1} = (-0.4,0.0)$, as obtained for a homogeneous superfluid constrained in a ring geometry (diamonds) and with the additional
presence of a weak point-like defect placed inside the ring (triangles).
In the experiment dissipation is expected to occur as soon as $\langle w_{F} \rangle < \langle w_{0} \rangle$.
However, while the onset of dissipation is observed in both geometries for the weaker coupling $(k_{F}a_{F})^{-1} = -0.4$ as shown in Fig.~\ref{Figure-7}(a),
at unitarity apparently no dissipation is found in either geometry as shown in Fig.~\ref{Figure-7}(b).
Here, we provide a theoretical explanation for this experimental finding in the following terms.

\begin{figure}[t]
\begin{center}
\includegraphics[width=7.8cm]{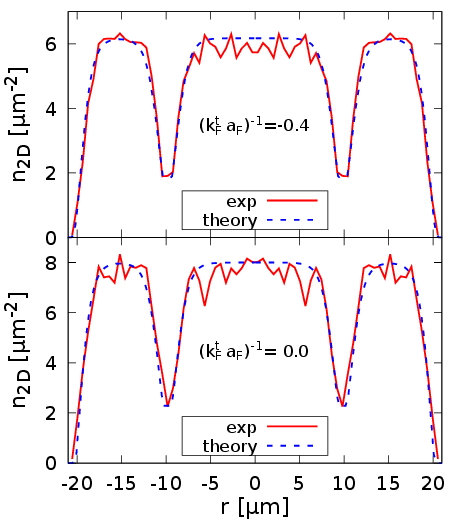}
\caption{Comparison between experimental (full lines) and theoretical (dashed  lines) density profiles with the ring geometry of Ref.~\cite{DelPace-2022} 
              for couplings $(k_{F}a_{F})^{-1} = -0.4$ (top panel) and $(k_{F}a_{F})^{-1} = 0.0$ (bottom panel).
              [Experimental profiles courtesy of G. Del Pace.]}
\label{Figure-8}
\end{center}
\end{figure}

We first unfold the ring into a linear tube of the same length and supplement it with periodic boundary conditions at its edges.
We then adopt a strategy similar to that recently utilized in Ref.~\cite{Piselli-2023} in terms of the mLPDA approach (see Appendix~\ref{sec:Appendix-A} below) and partition the tube into a large number of (961) tubular filaments,
each of which is treated as if it were a homogeneous superfluid with a given (local) density and a linear super-current flowing through it.  
To have full control of the density profiles spanning these filaments, we adjust the number of atoms as well as the height and width of the walls that contain the atomic cloud, in such a way to reproduce the corresponding experimental density profiles.
A comparison between the experimental and theoretical density profiles obtained in this way is shown in Fig.~\ref{Figure-8}  for the same couplings considered in 
Fig.~\ref{Figure-7}.

With this calibration procedure at hand, we may now restrict ourselves to considering the tubular filament that corresponds to the innermost part of the original ring with the smallest distance 
$R_{\mathrm{in}}$ from its center where the critical flow is reached first.
This is a consequence of the facts that all filaments must have the same winding number and that the quantization for the velocity obtained above reads $q = w /(2R)$ in terms of the linear momentum $q = v/m$ at
a distance $R$ from the ring center.
In this way, the intrinsic current density $j(q_{0})$ is computed for several momenta $q_{0} =w_{0}/(2R_{\mathrm{in}})$ along the lines of Sec.~\ref{pfchar}, and the
final winding number $w_{F}$ is obtained from the relations $q_{F}/m= j(q_0)/ n(R_{\mathrm{in}})$ and $q_{F} = w_{F} / (2R_{\mathrm{in}})$ where $n(R_{\mathrm{in}})$ is the local particle density at position
$R_{\mathrm{in}}$.
The results of this calculation are then reported in Fig.~\ref{Figure-7}, where they are compared with the experimental data of Ref.~\cite{DelPace-2022}.

Note from this figure that, as long as the momentum $q_{0}$ is below the critical momentum corresponding to the Landau dissipative threshold, the initial $w_{0}$ and final $w_{F}$ winding numbers coincide with each other as expected. 
Conversely, above this critical momentum (identified by the end point of the linear increase in Fig.~\ref{Figure-7}) the onset of the back-flow current acts to suppress the value of the super-current, thus making $w_{F} < w_{0}$.
Note also that the presence of a weak barrier on the experimental determination of the winding number has only a minor effect for coupling $(k_{F}a_{F})^{-1} = -0.4$ and essentially no effect for coupling $(k_{F}a_{F})^{-1} = 0.0$.
To confirm this experimental finding, we have also explicitly considered the presence of a weak barrier in our theoretical calculations, although only to realize that its effects are indeed quite negligible.
Note, finally, that Fig.~\ref{Figure-7}  predicts that, at unitarity, the theoretical value of the threshold at which $w_{F}$ deviates from a linear behavior considerably exceeds the maximum value of $w_{0}$ experimentally attainable.

\vspace{-0.2cm}
\section{Conclusions}
\label{sec:conclude}

In this article, we have performed a systematic investigation of the current-vs-momentum characteristics for a fermionic superfluid spanning the BCS-BEC crossover, for which the inclusion of pairing fluctuations beyond mean field plays a crucial role.
To this end, we have initially considered a fully homogeneous superfluid system, for which the above characteristics have allowed us to identify the \emph{intrinsic\/} upper value of the super-current 
flowing through the system, in terms of \emph{a single expression\/} of the super-current which is intrinsically limited by two different dissipative mechanisms (pair-breaking and phonon excitations) on the two BCS and BEC sides of the crossover, respectively.
In this context, we have considered both the Landau \cite{Landau-1941} and Bardeen \cite{Bardeen-1962} criteria for the loss of superfluid behavior when the intrinsic current exceeds a corresponding upper value,
and examined how their respective values evolve across the BCS-BEC crossover within suitable temperature ranges.
When needed, we have further included the effects of a small barrier that somewhat spoils the system homogeneity, so as to reproduce as closely as possible the experimental configurations based 
on linear \cite{Weimer-2015} and annular \cite{DelPace-2022} geometries.

In our analysis, we have not explicitly investigated the fate of the homogeneous superfluid once passed the Landau or Bardeen thresholds, when one expects patterns of inhomogeneous fluctuations to be built
from pair-breaking (on the BCS side) and sound (on the BEC side) elementary excitations, giving rise to  spatially non-uniform configurations with a strong local suppression of the gap parameter.
Typically, these configurations are expected to be precursors for the formation of vortices.  
This topic, although interesting in itself both theoretically and experimentally, exceeds the purposes of the present article.
In this respect, one may envisage that an appropriate theoretical method should possibly be based on a time-dependent approach, that would include pairing fluctuations and spatially inhomogeneous configurations on the same footing
over a wide range of temperature and coupling throughout the BCS-BEC crossover.

\vspace{0.2cm}
\noindent
\emph{Acknowledgments -} We are indebted to G. Roati and G. Del Pace for a critical reading of the manuscript.

\appendix 

\section{BRIEF OVERVIEW OF THE MAIN \\ PREVIOUS THEORETICAL RESULTS \\ UTILIZED IN THE PRESENT WORK}
\label{sec:Appendix-A}

\vspace{-0.1cm}
For the benefit of the readers, this Appendix briefly summarizes a number of theoretical results previously obtained in the literature, which are relevant to obtain the results discussed in the present work for the critical current 
with the inclusion of pairing fluctuations beyond mean field that span the BCS-BEC crossover.
The results here summarized include 
(i) the non-self-consistent $t$-matrix approach in the presence of a superfluid flow that was implemented in Ref.~\cite{Pisani-2023}, 
(ii) the modified Local Phase Density Approximation (mLPDA) that was also introduced in Ref.~\cite{Pisani-2023} to include pairing fluctuations over and above the original Local Phase Density Approximation (LPDA) of Ref.~\cite{Simonucci-2014}, 
and (iii) the extended Gorkov-Melik-Barkhudarov (GMB) approach originally introduced in Refs.~\cite{Pisani-2018-a} and \cite{Pisani-2018-b}.

\begin{center}
{\bf 1. Expressions for the density and current \\ in the presence of a superfluid flow}
\end{center}

In the superfluid phase, the fermionic local number density and current read
\begin{subequations}\label{eq:dens_cur_Green}
\begin{equation}\label{eq:density_Green}
n(\mathbf{r})=\dfrac{2}{\beta}\sum_ne^{i\omega_n\eta} G_{11}(\mathbf{r},\mathbf{r};\omega_n)
\end{equation}
\begin{equation}\label{eq:current_Green}
j(\mathbf{r})=\dfrac{1}{\beta}\sum_ne^{i\omega_n\eta}\dfrac{\left(\nabla_{\mathbf{r}}-\nabla_{\mathbf{r'}}\right)}{im} G_{11}(\mathbf{r},\mathbf{r'};\omega_n)\rvert_{\mathbf{r}=\mathbf{r'}} \, ,
\end{equation}
\end{subequations}
where $G_{11}$ is the ``normal" single-particle Green's function, $\beta = (k_{B}T)^{-1}$ the inverse temperature ($k_B$ being the Boltzmann constant), $\eta$ a positive infinitesimal, $m$ the fermion mass, and $\omega_n=(2n+1)\pi/\beta$ ($n$ integer) a fermionic Matsubara frequency \cite{Schrieffer-1964}.

When a super-current with momentum $\mathbf{q}$ flows in a homogeneous environment, the gap parameter takes the form \cite{DeGennes-1966}
\begin{equation}\label{eq:delta_q}
\Delta(\mathbf{r})=e^{i2\mathbf{q}\cdot\mathbf{r}}\Delta_\mathbf{q} \, .
\end{equation}
To comply with this spatial dependence, the single-particle Green's function $G_{11}$ then becomes \cite{Pisani-2023}
\begin{equation}\label{eq:trasf_11}
G_{11}(x,x';\mathbf{q})=e^{i\mathbf{q}\cdot(\mathbf{r}-\mathbf{r'})}\mathcal{G}_{11}(x-x';\mathbf{q})
\end{equation}
where $\mathcal{G}_{11}$ is the ``reduced'' single-particle Green's function
\begin{equation}\label{eq:trasf}
\mathcal{G}_{11}(x-x';\mathbf{q})=\sum_{k}e^{i\mathbf{k}\cdot(\mathbf{r}-\mathbf{r'})}e^{-i\omega_n(\tau-\tau')}\mathcal{G}_{11}(k;\mathbf{q})
\end{equation}
with the short-hand notation $\sum_k=\int\dfrac{d\mathbf{k}}{(2\pi)^3}\dfrac{1}{\beta}\sum_n$, $k=(\mathbf{k},\omega_n)$ being a fermionic four-vector.
Accordingly, in momentum and frequency space the expressions \eqref{eq:density_Green} and \eqref{eq:current_Green} read
\begin{subequations}\label{eq:nj}
\begin{equation}
n=2 \int \frac{d \mathbf{k}}{(2 \pi)^3} \frac{1}{\beta} \sum_n e^{i \omega_n \eta} \mathcal{G}_{11}\left(\mathbf{k}, \omega_n ; \mathbf{q}\right)
\end{equation}
\begin{equation}\label{eq:j}
\mathbf{j}=\frac{\mathbf{q}}{m} n+2 \int \frac{d \mathbf{k}}{(2 \pi)^3} \frac{\mathbf{k}}{m} \frac{1}{\beta} \sum_n e^{i \omega_n \eta} \mathcal{G}_{11}\left(\mathbf{k}, \omega_n ; \mathbf{q}\right) \, .
\end{equation}
\end{subequations}

The form of $\mathcal{G}_{11}\left(k; \mathbf{q}\right)$ to be entered in the above expressions can be evaluated within different approximations, depending on the choice of the single-particle self-energy. 
In the present work, it is evaluated both at the mean-field level and with the inclusion of pairing fluctuations within the non-self-consistent $t$-matrix approach.
Quite generally, $\mathcal{G}_{11}$ is obtained by solving the Dyson's equation in the broken-symmetry phase in the Nambu-Gorkov formalism \cite{Pisani-2023}
\begin{widetext}
\begin{equation}\label{eq:Dyson}
\begin{bmatrix}
i\omega_n-\xi(\mathbf{k}+\mathbf{q})-\mathfrak{S}_{11}(k;\mathbf{q}) & -\mathfrak{S}_{12}(k;\mathbf{q}) \\
-\mathfrak{S}_{21}(k;\mathbf{q}) & i\omega_n+\xi(\mathbf{k}-\mathbf{q})-\mathfrak{S}_{22}(k;\mathbf{q})
\end{bmatrix}
\begin{bmatrix}
\mathcal{G}_{11}(k;\mathbf{q}) & \mathcal{G}_{12}(k;\mathbf{q}) \\
\mathcal{G}_{21}(k;\mathbf{q}) & \mathcal{G}_{22}(k;\mathbf{q})
\end{bmatrix}
=
\begin{bmatrix}
1 & 0 \\
0 & 1 
\end{bmatrix} \, ,
\end{equation} 
\end{widetext}
where $\xi(\mathbf{k})=\mathbf{k}^2/(2m)-\mu$ and $\mathfrak{S}_{ii'}$ are the components of the ``reduced'' self-energy.
Different approximations then correspond to different choices of $\mathfrak{S}_{ii'}$.

\begin{center}
{\bf 1-a - Mean-field approximation \\ in the presence of a super-current}
\end{center}

At the mean-field ($\mathrm{mf}$) level, one takes $\mathfrak{S}_{11}(k,\mathbf{q})=\mathfrak{S}_{22}(k,\mathbf{q})=0$ and $\mathfrak{S}_{12}(k,\mathbf{q})=\mathfrak{S}_{21}(k,\mathbf{q})=-\Delta_\mathbf{q}$,
such that solving the Dyson's equation \eqref{eq:Dyson} yields \cite{Pisani-2023}
\begin{subequations}\label{eq:Green_mf}
\begin{equation}\label{eq:g11}
\mathcal{G}_{11}^{\mathrm{mf}}(k;\mathbf{q})=\dfrac{u(\mathbf{k};\mathbf{q})^2}{i\omega_n-E_+(\mathbf{k};\mathbf{q})}+\dfrac{v(\mathbf{k};\mathbf{q})^2}{i\omega_n+E_-(\mathbf{k};\mathbf{q})}
\end{equation}
\begin{equation}
\begin{split}
\mathcal{G}_{12}^{\mathrm{mf}}(k;\mathbf{q})&=-u(k;\mathbf{q})v(k;\mathbf{q}) \\
&\times \left[\dfrac{1}{i\omega_n-E_+(\mathbf{k};\mathbf{q})}
-\dfrac{1}{i\omega_n+E_-(\mathbf{k};\mathbf{q})}\right],
\end{split}
\end{equation}
\end{subequations}
where
\begin{subequations}
\begin{equation}
u(\mathbf{k};\mathbf{q})^2=\dfrac{1}{2}\left(1+\dfrac{\xi(\mathbf{k};\mathbf{q})}{E(\mathbf{k};\mathbf{q})}\right)
\end{equation}
\begin{equation}
v(\mathbf{k};\mathbf{q})^2=\dfrac{1}{2}\left(1-\dfrac{\xi(\mathbf{k};\mathbf{q})}{E(\mathbf{k};\mathbf{q})}\right)
\end{equation}
\end{subequations}
with the notation
\begin{equation}
\begin{split}
\xi(\mathbf{k};\mathbf{q})&=\dfrac{\mathbf{k}^2}{2m}-\mu+\dfrac{\mathbf{q}^2}{2m} \\
E(\mathbf{k};\mathbf{q})&=\sqrt{\xi(\mathbf{k};\mathbf{q})^2+\Delta_{\mathbf{q}}^2} \\
E_{\pm}(\mathbf{k};\mathbf{q})&=E(\mathbf{k};\mathbf{q})\pm\dfrac{\mathbf{k}\cdot\mathbf{q}}{m} \, .
\label{useful-notation}
\end{split}
\end{equation}
Entering the result (\ref{eq:g11}) for $\mathcal{G}_{11}$ into the expressions (\ref{eq:nj}), one ends up with the expression (\ref{bcslandau}) for the current density at the mean-filed level
(as well as with the associated expression for the density), which has the typical form of the two-fluid model at finite temperature in the Bardeen formulation for fermions \cite{Bardeen-1962}.

\begin{center}
{\bf 1-b - $t$-matrix approximation \\ in the presence of a super-current}
\end{center}

\vspace{-0.2cm}
Within the pairing-fluctuation ($\mathrm{pf}$) approximation, the reduced self-energy of the Dyson's equation (\ref{eq:Dyson}) reads \cite{Pisani-2023}
\begin{align}
	\left\{\begin{array}{l}\mathfrak{S}_{11}^\mathrm{pf}(k ; \vq)=-\mathfrak{S}_{22}^\mathrm{pf}(-k ; \vq) \\
	\hspace{1.55cm} = - \sum_Q \Gamma_{11}(Q ; \vq) \mathcal{G}_{11}^\mathrm{mf}(Q-k ; \vq) \\ 
	\\
	\mathfrak{S}_{12}^\mathrm{pf}(k ; q)=\mathfrak{S}_{21}^\mathrm{pf}(k ; q)=-\Delta_\vq                  \end{array}  \right.   
	\label{se}
\end{align}
with the short-hand notation $\sum_Q \longleftrightarrow \int \frac{d \vQ}{(2 \pi)^3} \frac{1}{\beta} \sum_\nu$,
where $Q=\left(\vQ, \Omega_v\right)$ is a four-vector with $\Omega_\nu=2 \nu \pi / \beta$ ($\nu$ integer) a bosonic Matsubara frequency \cite{FW-1971}.
In this way, the expression of $\mathcal{G}_{11}\left(k; \mathbf{q}\right)$, to be utilized in Eq.~(\ref{eq:j}) to obtain the current with the inclusion of pairing fluctuations, becomes:
\begin{widetext}
\begin{equation}\label{eq:g11_pf}
\mathcal{G}_{11}^{\mathrm{pf}}(k ; \mathbf{q})=\frac{1}{i \omega_n-\xi(\mathbf{k}+\mathbf{q})-\mathfrak{S}_{11}^{\mathrm{pf}}(k ; \mathbf{q})-\frac{\Delta_{\mathbf{q}}^2}{i \omega_n+\xi(\mathbf{k}-\mathbf{q})+\mathfrak{S}_{11}^{\mathrm{pf}}(-k ; \mathbf{q})}} \, .
\end{equation}
\end{widetext}

The quantity $\Gamma_{11}$ entering Eq.~(\ref{se}) is the upper diagonal element of the $2 \times 2$ matrix for the ``pair propagator'', which consists in a series of ladder diagrams whereby two fermions with opposite spins repeatedly scatter with each other
\cite{Andrenacci-2003}.
In the presence of a super-current, its expression reads \cite{Pisani-2023}
\begin{align}\label{eq:gamma_11}
{\left[\begin{array}{ll}
\Gamma_{11}(Q ; \vq) & \Gamma_{12}(Q ; \vq) \\
\Gamma_{21}(Q ; \vq) & \Gamma_{22}(Q ; \vq) 
\end{array}\right] } & =\frac{1}{A(Q ; \vq) A(-Q ; \vq)-B(Q ; \vq)^2} \nonumber \\
& \times\left[\begin{array}{cc}
A(-Q ; \vq) & B(Q ; \vq) \\
B(Q ; \vq) & A(Q ; \vq)
\end{array}\right]
\end{align}
where
\begin{align}
\mathrm{A}(Q ; \mathbf{q}) & =-\frac{m}{4 \pi a_F}+\int \frac{d \mathbf{k}}{(2 \pi)^3} \frac{m}{\mathbf{k}^2} \nonumber \\
	&  - \sum_k \mathcal{G}_{11}^{\mathrm{mf}}(k+Q ; \mathbf{q}) \mathcal{G}_{11}^{\mathrm{mf}}(-k ; \mathbf{q})  \label{Aq}\\
	B(Q ; \mathbf{q}) & =\sum_k \mathcal{G}_{12}^{\mathrm{mf}}(k+Q ; \mathbf{q}) \mathcal{G}_{12}^{\mathrm{mf}}(-k ; \mathbf{q})  \label{Bq}
\end{align}
with the normal $\mathcal{G}_{11}^\mathrm{mf}(k;\vq)$ and anomalous $\mathcal{G}_{12}^\mathrm{mf}(k;\vq)$ mean-field  Green's functions given by Eqs.~(\ref{eq:Green_mf}).
Performing the sums over the Matsubara frequency in Eqs.~(\ref{Aq}) and (\ref{Bq}), one obtains for the particle-particle rungs the rather lengthy expressions reported in detail in Appendix A of Ref.~\cite{Pisani-2023},
which need not be reported here.
It may be instead important to emphasize that the presence of the pair propagator (\ref{eq:gamma_11}) in the $t$-matrix approach guarantees that the effects not only of pair-breaking (single-particle) excitations but also of sound-mode (two-particle) excitations are present in the physical quantities calculated in terms of this approach.

In addition, in Appendix B of Ref.~\cite{Pisani-2023} it was shown that, when approximating the quantities $A(Q ; \vq)$ and $B(Q ; \vq)$ in the BEC limit of the BCS-BEC crossover, the expression of the current, obtained by utilizing 
$\mathcal{G}_{11}^{\mathrm{pf}}(k ; \mathbf{q})$ of Eq.~(\ref{eq:g11_pf}) into Eq.~(\ref{eq:j}), recovers the typical form of the current within a two-fluid model \cite{PS-2008} for a bosonic gas treated with the Bogoliubov approximation.
This form, in turn, coincides with the expression (\ref{beclandau}) considered in Sec.~\ref{sec:wrbg}.

\begin{center}
{\bf 2. The mLPDA approach}
\end{center}

The mLPDA approach introduced in Ref.~\cite{Pisani-2023} consists of a ``modified'' Local Phase Density Approximation approach, in which the inclusion of pairing fluctuations at the level of the non-self-consistent $t$-matrix was implemented 
on top of the original LPDA approach of Ref.~\cite{Simonucci-2014}. 
In particular, the LPDA approach was utilized to study the Josephson effect occurring when a barrier is embedded in a homogeneous fermionic superfluid spanning the BCS-BEC crossover \cite{Piselli-2020}, 
a problem for which the mLPDA approach was further considered \cite{Piselli-2023} aiming at comparing with recent experimental measurements of the Josephson critical current in ultra-cold atomic Fermi gases \cite{Kwon-2020,DelPace-2021}.

When a super-current flows across a barrier embedded in an otherwise homogeneous superfluid, the order parameter reads
\begin{equation}\label{eq:delta_LPDA}
\Delta(x)=|\tilde{\Delta}(x)| e^{2i\mathbf{q}\cdot\mathbf{x}+2i\phi(x)}=e^{2i\mathbf{q}\cdot\mathbf{x}}\tilde{\Delta}(x) \, ,
\end{equation}
\noindent
where $\hat{\mathbf{x}}$ is the direction of the superfluid flow and $2\phi(x)$ the phase of the order parameter due to the presence of the barrier, in addition to that occurring in Eq.~(\ref{eq:delta_q}).

In this case, the LPDA equation takes the form \cite{Piselli-2020}
\begin{equation}\label{eq:LPDA}
\begin{split}
-\dfrac{m}{4\pi a_F}\tilde\Delta(x)&=\mathcal{I}_0(\mathbf{x})\tilde\Delta(x)+\dfrac{\mathcal{I}_1(x)}{4m}\dfrac{\text{d}^2}{\text{d}x^2}\tilde\Delta(x)\\
&+i\mathcal{I}_1(x)\dfrac{q}{m}\dfrac{\text{d}\tilde{\Delta}(x)}{\text{d}x},
\end{split}
\end{equation}
\noindent
where the coefficients $\mathcal{I}_0$ and $\mathcal{I}_1$ are given by
\begin{subequations}\label{eq:I_0I_1}
\begin{equation}\label{eq:I_0}
\mathcal{I}_0(x)=\int\!\!\dfrac{\text{d}\mathbf{k}}{(2\pi)^3}\left[\dfrac{1-2f_F(E_+^\mathbf{q}(\mathbf{k}|x))}{2E(\mathbf{k}|x)}-\dfrac{m}{\mathbf{k}^2}\right]
\end{equation}
\begin{equation}\label{eq:I_1}
\begin{split}
\mathcal{I}_1(x)&=\dfrac{1}{2}\int\!\!\dfrac{\text{d}\mathbf{k}}{(2\pi)^3}\bigg\{\dfrac{\xi(\mathbf{k}|x)}{2E(\mathbf{k}|x)^3}[1-2f_F(E_+^\mathbf{q}(\mathbf{k}|x))]\\
&+\dfrac{\xi(\mathbf{k}|x)}{2E(\mathbf{k}|x)^2}\dfrac{\partial f_F(E_+^\mathbf{q}(\mathbf{k}|x))}{\partial E_+^\mathbf{q}(\mathbf{k}|x)}\\
&+\dfrac{\mathbf{k}\cdot\mathbf{q}}{\mathbf{q}^2}\dfrac{1}{E(\mathbf{k}|x)}\dfrac{\partial f_F(E_+^\mathbf{Q_0}(\mathbf{k}|x))}{\partial E_+^\mathbf{q}(\mathbf{k}|x)}\bigg\}
\end{split}
\end{equation}
\end{subequations}
\noindent
with
\begin{equation}\label{eq:xiEE+}
\begin{split}
\xi(\mathbf{k}|x)&=\dfrac{\mathbf{k}^2}{2m}-\left[\mu -V_{\mathrm{ext}}(x)-\dfrac{\mathbf{q}^2}{2m}\right] \\
E(\mathbf{k}|x)&=\sqrt{\xi(\mathbf{k}|x)^2+|\tilde{\Delta}(x)|^2} \\ 
E_+^\mathbf{q}(\mathbf{k}|x)&=E(\mathbf{k}|x)+\dfrac{\mathbf{k}\cdot\mathbf{q}}{m} \, ,
\end{split}
\end{equation}
\noindent
$V_{\mathrm{ext}}(x)$ being the external potential associated with the Josephson barrier.

When implementing the numerical calculations, the imaginary part of the LPDA equation (\ref{eq:LPDA}) is conveniently replaced by the constraint of the current conservation \cite{Piselli-2020,Pisani-2023,Piselli-2023}, namely,
\begin{equation}
j(x) - J=0 \, ,
\label{current-conservation}
\end{equation}
where $J$ is the current evaluated far from the barrier. 

Within the LPDA approach of Ref.~\cite{Simonucci-2014}, the expression for the local current $j(x)$ to be utilized in Eq.~(\ref{current-conservation}) is obtained by entering the mean-filed form (\ref{eq:g11}) for $\mathcal{G}_{11}\left(k; \mathbf{q}\right)$ 
into the expression (\ref{eq:j}) for the current, provided the following local replacements are performed therein:
\begin{subequations}
\begin{equation}\label{eq:locmu}
\mu\longrightarrow\mu-V_{\mathrm{ext}}(x)
\end{equation}
\begin{equation}
\Delta_{\mathbf{q}}\longrightarrow|\Delta(x)|
\end{equation}
\begin{equation}\label{eq:trasf_q}
\mathbf{q}\longrightarrow\mathbf{q}+\dfrac{d\phi(x)}{dx} \, .
\end{equation}
\end{subequations}

Within the mLPDA approach, on the other hand, the local replacement \eqref{eq:locmu} may lead to unwanted unphysical singularities in the diagonal element $\Gamma_{11}(Q;\mathbf{q})$ \eqref{eq:gamma_11} of the pair propagator, 
when this is treated within a local perspective with local values of the gap parameter and of the chemical potential.
For this reason, in Ref.~\cite{Pisani-2023} it was found convenient to utilize the local requirement 
\begin{equation}
\mu\longrightarrow\mu-V_{\mathrm{eff}}(x)
\end{equation}
in the place of Eq.~(\ref{eq:locmu}), where now $V_{\mathrm{eff}}(x)$ is a suitable ``effective'' potential which ensures the gapless condition at  $Q=0$ of the pair propagator at any $x$.
Examples of the spatial profile of $V_{\mathrm{eff}}(x)$ were reported in Fig.~2 of Ref.~\cite{Pisani-2023} for several couplings and temperatures, when the external potential $V_{\mathrm{eff}}(x)$ has a Gaussian form.

\begin{center}
{\bf 3. The extended Gorkov-Melik-Barkhudarov approach}
\end{center}

In Ref.~\cite{Pisani-2018-b} a diagrammatic scheme was implemented for improving on the treatment of pairing fluctuations over and above the $t$-matrix approach, 
by generalizing to the whole BCS-BEC crossover the original work by Gor'kov and Melik-Barkhudarov (GMB) which was meant for the BCS limit only \cite{GMB-1961}.
To this end, the property of the pair propagator $\mathbf{\Gamma}(\vQ,\Omega)$ to be equivalent to the Bogoliubov propagators for composite bosons that form in the BEC limit of the BCS-BEC crossover \cite{Andrenacci-2003}
was first utilized to show that, already at the level of the $t$-matrix approach in the superfluid phase, the gap equation for the constituent fermions throughout the whole BCS-BEC crossover is equivalent to a bosonic Hugenholtz-Pines condition, 
in the form $\Gamma_{11}^{-1}(0,0)-\Gamma^{-1}_{12}(0,0)=0$.
In this way, it was possible to go beyond the $t$-matrix approach in a natural way, by introducing a suitable bosonic-like self-energy correction $\mathbf{\Sigma_B}(\vQ,\Omega_{\nu})$ to the bare pair propagator $\mathbf{\Gamma}(\vQ,\Omega_{\nu})$,
such that 
\begin{eqnarray}
\mathbf{\Gamma}^{-1}(\vQ,\Omega_{\nu}) & \rightarrow & \mathbf{\Gamma}_{\mathrm{dressed}}^{-1}(\vQ,\Omega_{\nu}) 
\nonumber \\
& = & \mathbf{\Gamma}^{-1}(\vQ,\Omega_{\nu}) - \mathbf{\Sigma_B}(\vQ,\Omega_{\nu})
\label{dyst}
\end{eqnarray} 
in a way formally equivalent to a Dyson's equation.
In addition, in Ref.~\cite{Pisani-2018-b} the fermionic self-energy was kept of the $t$-matrix form (that is, like in Eq.~(\ref{se})), with the bare pair propagator replaced, however, by the dressed one of Eq.~(\ref{dyst}). 

As mentioned above, the form of the bosonic-like self-energy $\mathbf{\Sigma_B}$ adopted in Ref.~\cite{Pisani-2018-b} was motivated by the original work by Gor'kov and Melik-Barkhudarov \cite{GMB-1961},
who considered the screening of the pairing interaction due to the polarization of the surrounding medium, as represented diagrammatically at second order in the inter-particle interaction by two crossing interaction lines in a particle-hole rung. 
This effect was shown to introduce a correction by a factor of $(4e)^{1/3}$ to the BCS values of both the critical temperature and the pairing gap \cite{GMB-1961}.
In Ref.~\cite{Pisani-2018-b} the GMB correction of Ref.~\cite{GMB-1961} was \emph{extended\/} to the whole BCS-BEC crossover, both in the normal and superfluid phases, by identifying the diagram representing the bosonic-like self-energy to be inserted
in Eq.~(\ref{dyst}) with the original GMB diagram, where the interaction lines of the particle-hole rung are now replaced by pair propagators $\mathbf{\Gamma}(\vQ,\Omega)$ which include
infinitely repeated scattering between two fermions and thus go beyond second order (cf. Fig.~2 of Ref.~\cite{Pisani-2018-b}).

In Ref.~\cite{Pisani-2018-b} the fermionic superfluid was considered at rest, which corresponds to the case with $\mathbf{q} = 0$ in Eqs.~(\ref{se})-(\ref{Bq}).
Full consideration of the GMB correction in the presence of a superfluid flow with $\mathbf{q} \ne 0$ is beyond the scope of the present work.
Nonetheless, we may still consider the effect of the GMB correction in an approximate manner, by replacing the bare pair propagator $\mathbf{\Gamma}$ of Eq.~(\ref{dyst}) by the dressed one $\mathbf{\Gamma}_{\mathrm{dressed}}$ of Eq.~(\ref{dyst}),
where now $\mathbf{\Gamma}(\vQ,\Omega_{\nu}) \rightarrow \mathbf{\Gamma}(\vQ,\Omega_{\nu};\vq)$ contains $\mathbf{q} \ne 0$ while the bosonic-like self-energy correction $\mathbf{\Sigma_B}$ is taken from the expression with
$\mathbf{q} = 0$ given in Ref.~\cite{Pisani-2018-b} in the absence of current, where we further set $\vQ=0$ and $\Omega_{\nu}=0$.
As already noted in Ref.~\cite{Pisani-2018-b}, at low temperature the effect of the GMB self-energy amounts in practice to a coupling-dependent shift of the coupling strength $(k_{F} a_{F})^{-1}$ entering in the bare inverse propagator 
$\mathbf{\Gamma}^{-1}(\vQ,\Omega_{\nu})$.
This remark allows for a swift implementation of the GMB correction in the present work, as it was done in Sec.~\ref{sec:intr}-C.

\vspace{-0.4cm}
\section{ADDENDUM ABOUT TWO TOPICS TREATED IN THE MAIN TEXT}
\label{sec:Appendix-B}

\vspace{-0.1cm}
This Appendix expands on two topics dealt with in Secs.~\ref{sec:bcsj} and \ref{sec:LB-intrinsic}, regarding respectively:
(i) The free energy of a weakly-attractive Fermi gas treated at the mean-field level, thus illustrating in more detail the Bardeen criterion for the supercritical flow (given by Eq.~(\ref{bardcrit}) of Sec.~\ref{sec:bcsj}) 
in terms of an analytically solvable approximation;
(ii) The crossing between ascending and descending branches of two-particle excitations, which is shown to occur on the BEC side of unitarity for a coupling at which the underlying Fermi surface has not yet collapsed.

\begin{center}
{\bf 1. Free energy at the mean-field level in the presence of a super-current}
\end{center}

\begin{figure}[t]
\begin{center}
\includegraphics[width=8.2cm]{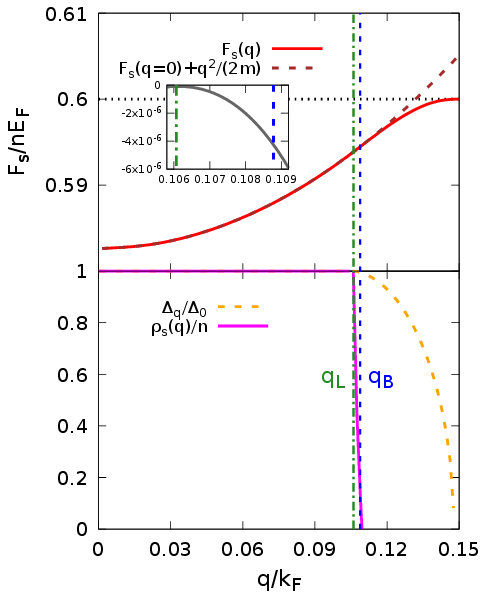}
\caption{Upper panel: The free energy density $F_{s}(q)$ given by Eq.~(\ref{free-energy-density}) (red full line) is shown vs the magnitude $q$ of the momentum (in units of the Fermi wave vector $k_{F}$) for coupling $(k_{F} a_{F})^{-1} = -1.0$,
                                   where it is also compared with the quantity $F_{s}(q=0)+\frac{q^{2}}{2m}$ (brown dashed line) and with the free energy density $F_{n}={3 \over 5} n E_{F}$ of an ideal Fermi gas (horizontal black dotted line).
                                   Also shown are the Landau critical value $q_{\mathrm{L}}$ (vertical green dashed-dotted line) and the Bardeen critical value $q_{\mathrm{B}}$ (vertical blue dashed line).
                                   The inset shows the difference $F_{s}(q=0)+\frac{q^{2}}{2m} - F_{s}(q)$ in the range between (about) $q_{\mathrm{L}}$ and $q_{\mathrm{B}}$.
              Lower panel: The corresponding superfluid order parameter $\Delta_{q}$ (orange dashed line) and superfluid density $\rho_{s}(q)$ (violet full line) are shown vs $q$.}
\label{Figure-9}
\end{center}
\end{figure}

The free energy density $F_{s}(\vq)$ of a weakly-attractive Fermi gas treated at the mean-field level can be computed in the superfluid phase at the mean-field level by following Ref.~\cite{Pistolesi-1996}, 
provided that the Nambu Green's functions in Eq.~(2.35) therein are replaced by their counterpart (\ref{eq:Green_mf}) in the presence of a super-current.
In this way, one obtains the following expression
\begin{eqnarray}
F_{s}(\vq) & = & - { \Delta_q^2 \over v_{0} } +\int \frac{d\vk}{(2\pi)^3} \bigg( \xi(\vk,\vq)-E(\vk,\vq) \bigg) 
\label{free-energy-density} \\
& - &{ 2 \over \beta } \int \frac{d\vk}{(2\pi)^3} \log \left[1 + \exp\bigg(\!\!\! -\beta E_+(\vk,\vq) \! \bigg) \right] +\mu_{\vq} n
\nonumber 
\end{eqnarray}
with the notation (\ref{useful-notation}) (cf. also Eq.~(42) of Ref.~\cite{Taylor-2006}). 
In the above expression, $v_{0} < 0$ is the strength of the contact inter-particle interaction and $\mu_{\vq}$ is the chemical potential in the presence of a super-current which depends on $q=|\vq|$ when $q > q_{\mathrm{L}}$.

Limiting to zero temperature, the quantity (\ref{free-energy-density}) for coupling $(k_{F} a_{F})^{-1} = -1.0$ is shown vs $q = |\vq|$ in the upper panel of  Fig.~\ref{Figure-9}  (red full line), 
where it is compared with its counterpart in the normal state $F_{n}={3 \over 5} n E_{F}$ with Fermi energy $E_{F} = \frac{k_{F}^{2}}{2m}$ (horizontal black dotted line) which represents the free energy density of an ideal Fermi gas.
This panel shows also the quantity $F_{s}(q=0)+\frac{q^{2}}{2m}$ (brown dashed line), which benchmarks the superfluid non-dissipative state and thus coincides with $F_{s}(\vq)$ up to the Landau critical value $q_{\mathrm{L}}$
(vertical green dashed-dotted line) at which dissipation sets in. 
Also shown is the Bardeen critical value $q_{\mathrm{B}}$ (vertical blue dashed line) in correspondence with the inset of  Fig.~\ref{Figure-1}  of the main text which shows $j(q)=\partial F_s(q)/ \partial q$. 
In the inset of  Fig.~\ref{Figure-9}, the difference $F_{s}(q=0)+\frac{q^{2}}{2m} - F_{s}(q)$ is plotted in a restricted range between about $q_{\mathrm{L}}$ and $q_{\mathrm{B}}$, 
where this quantity starts being different from zero at $q_{\mathrm{L}}$. 

In the lower panel of  Fig.~\ref{Figure-9}  the superfluid order parameter $\Delta_{q}$ of Eq.~(\ref{eq:delta_q}) (orange dashed line) is shown vs $q$, where it is seen to vanish way beyond the Bardeen critical value $q_{\mathrm{B}}$.  
Also shown is the superfluid density $\rho_{s}(q) = \partial F_{s}^2/\partial q^2$ (violet full line) given by Eq.~(\ref{superfluid-density}) of the main text, which is seen to coincide with the full particle density $n$ up to the Landau critical value $q_{\mathrm{L}}$,
after which it abruptly decreases and eventually vanishes at the Bardeen critical value $q_{\mathrm{B}}$.
At this point, where the superfluid density $\rho_{s}$ would change sign, the thermodynamical stability of the system is lost.
Since $\rho_{s}$ represents the stiffness (or rigidity) of the superfluid component to a perturbing velocity field, its turning negative signals the instability of the translational state of the system,
in analogy with the occurrence of a negative compressibility in an unstable mechanical system (like for the liquid-gas transition).
The system thus undergoes a phase transition to a new state, which in the present case would most likely involve the condensation of finite-momentum Cooper pairs.

\begin{center}
{\bf 2. Crossing between ascending and descending branches of two-particle excitations}
\end{center}

One may wonder whether there exists a connection, between the coupling at which the crossover from pair-breaking to phonon excitations occurs (which determines the maximum current in Figs.~\ref{Figure-4} and \ref{Figure-5}) 
and the coupling at which the underlying Fermi surface collapses (such that at this ``splitting point'' the single-particle dispersion switches over from BCS-type to bosonic-type \cite{Son-2006}).
The detailed analysis reported below shows that this is actually not the case, such that the two couplings (albeit close in values) are physically unrelated to each other.

\begin{figure}[t]
\begin{center}
\includegraphics[width=9.0cm]{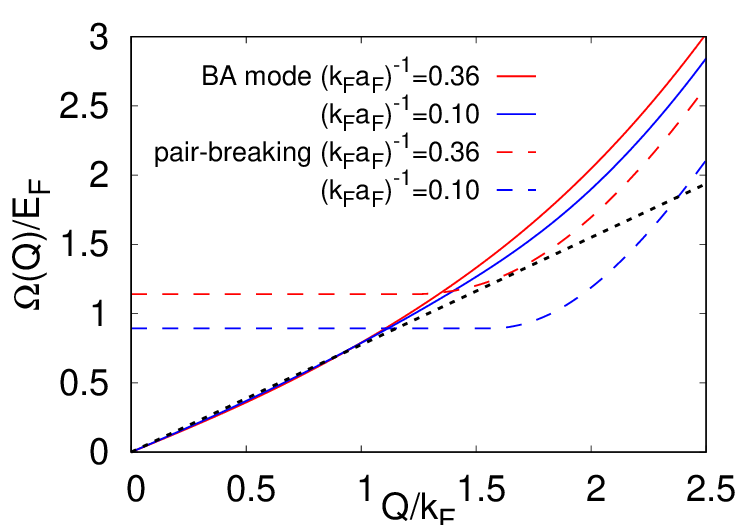}
\caption{Two-particle excitation spectrum for the crossover coupling $(k_{F}a_{F})^{-1} = + 0.36$ (red curves) and the weaker coupling $(k_{F}a_{F})^{-1} = +0.10$ (blue curves), where the solid lines correspond to 
            the sound mode and the dashed lines to the onset of the pair-breaking spectrum at zero temperature. 
            The black dotted line represents the Landau critical velocity for $(k_{F}a_{F})^{-1} = +0.36$, where this slope is identical for both pair-breaking and sound mode excitations.
            [Here, the wave vector $Q$ is in units of the Fermi wave vector $k_{F}$ and the frequency $\Omega$ in units of the Fermi energy $E_{F}$.]}
\label{Figure-10}
\end{center}
\end{figure}

In our calculation, the maximum of the Landau velocity equals $0.33v_{F}$ obtained for the coupling $(k_{F}a_{F})^{-1} = +0.36$ (cf. the red full line in Fig.~\ref{Figure-4}). 
This coupling corresponds to the point where the pair-breaking and sound velocities are comparable to each other (cf. the blue dotted line and the black dashed-dotted line in Fig.~\ref{Figure-4}), 
such that the two types of excitations have comparable energy. 
To better illustrate how the change between one type (pair-breaking) to the other type (sound mode) of excitations comes about, following what was done in Fig.~7 of Ref.~\cite{Pieri-2004}, Fig.~\ref{Figure-10} shows the zero-temperature
\emph{two-particle excitation spectrum\/} both for the crossover coupling $(k_{F}a_{F})^{-1} = + 0.36$ (red curves) and for the weaker coupling $(k_{F}a_{F})^{-1} = +0.10$ (blue curves), 
where the solid lines correspond to the sound mode and the dashed lines to the onset of the pair-breaking spectrum. 
The latter quantity is obtained like in Ref.~\cite{Pieri-2004}, but now considering the renormalized chemical potential $\mu_{\mathrm{L}} = k_{\mathrm{L}}^{2}/(2m)$ (where $k_{\mathrm{L}}$ is the so-called ``Luttinger'' wave vector to be discussed below)
and the excitation gap $\Delta_{\mathrm{e}}$ as extracted from the single-particle excitation spectrum, rather than the thermodynamic values $\mu$ and  $\Delta$ utilized in Ref.~\cite{Pieri-2004}
(including in this way some degree of self-consistency in the calculation which takes into account the effect of quantum fluctuations.
In Fig.~\ref{Figure-10} the slope of the black dotted line represents the Landau critical velocity for the crossover coupling $(k_{F}a_{F})^{-1} = +0.36$, where this slope is identical for both pair-breaking and sound mode excitations. 
For the weaker coupling $(k_{F}a_{F})^{-1} = +0.10$, on the other hand, the slope of the sound mode is evidently larger than that of the onset of the pair-breaking excitations, consistently with what is reported in Fig.~\ref{Figure-4}.

Regarding instead the coupling at which the collapse of the underlying Fermi surface takes place upon approaching the BEC side of the BCS-BEC crossover, this coupling refers to a property of the \emph{single-particle excitation spectrum\/} 
of the Fermi system and has been identified as the coupling at which the so-called Luttinger wave vector $k_{L}$ mentioned above vanishes.
In particular, this quantity was determined at the critical temperature $T_{c}$ by including pairing fluctuations beyond mean field within the $t$-matrix approximation, either when comparing with the available experimental data in ultra-cold Fermi gases \cite{Perali-2011}, or when considering in details the properties of the single-particle spectral function in the normal phase of a Fermi gas \cite{Palestini-2012}, and even with the inclusion of disorder \cite{Palest-2013}.
These calculations show that the splitting point ``$\kappa_{0}$'' of Ref.~\cite{Son-2006}, which is the coupling where $k_{L}$ vanishes according to Refs.~\cite{Perali-2011,Palestini-2012,Palestini-2013}, is about $+0.60$. 
This confirms our expectation that the crossover coupling $(k_Fa_F)^{-1} = +0.36$ (for two-particle excitations) and the splitting point  (for single-particle excitations) are two unrelated quantities.

	

\end{document}